\documentclass[onecolumn,sort&compress,numbers]{els-mrw} 

\usepackage{amsmath,amssymb,amsfonts,amsthm,makeidx,graphicx}
\usepackage{txfonts}
\usepackage{helvet}
\usepackage{soul}
\usepackage{xspace}

\newcommand{\eqn}{equation}
\newcommand{\lb}{\left(}
\newcommand{\rb}{\right)}
\newcommand{\GeV}{\ensuremath{\mathrm{GeV}}\xspace}
\newcommand{\TeV}{\ensuremath{\mathrm{TeV}}\xspace}

\newcommand{\be}{\beta}
\newcommand{\al}{\alpha}

\begin{document}



\chapter{BSM: Extended Scalar Sectors}\label{chap:bsmscalars}

\author[1,2]{Tania Robens}%
\author[3,4]{Rui Santos}%


\address[1]{\orgname{Rudjer Boskovic Institute}, \orgdiv{Division of Theoretical Physics}, \orgaddress{Bijenicka cesta 54, 10000 Zagreb, Croatia}}
\address[2]{\orgname{CERN}, \orgdiv{Theoretical Physics Department}, \orgaddress{1211 Geneva 23, Switzerland}}

\address[3]{\orgname{ISEL – Instituto Superior de Engenharia de Lisboa,
Instituto Polit\'ecnico de Lisboa}, \orgdiv{Departamento de Física}, \orgaddress{R. Conselheiro Emídio Navarro 1,  1959-007 Lisboa, Portugal}}
\address[4]{\orgname{Centro de F\'isica T\'eorica e Computacional, Faculdade de C\^iencias,
Universidade de Lisboa}, \orgdiv{Departamento de Física}, \orgaddress{Campo Grande, Edif\'icio C8 1749-016 Lisboa, Portugal}}

\articletag{Chapter Article tagline: update of previous edition, reprint.}

\maketitle

\begin{abstract}[Abstract]

    {
    In particle physics the world is described by a function, the Lagrangian. Each of its sectors characterizes the interactions between the particles of the Standard Model (SM). The addition of hypothetical new particles 
    is done by including new terms in the Lagrangian. 
    The scalar or Higgs sector of the SM is built with only one scalar complex field and it is extended by including new spin zero fields. 
    This can help to solve questions that cannot be answered by the SM alone, like introducing dark matter candidates or new sources of CP-violation required to explain the matter-antimatter asymmetry of the universe.
    The corresponding theories have to be probed experimentally. For the high energy region, the standard tools are collider experiments such as the Large Hadron Collider, or other possible future facilities.
    Dark matter experiments scrutinize the connection between the visible and the dark world.  \\
    RBI-ThPhys-2025-031, CERN-TH-2025-148

    }
\end{abstract}

\begin{keywords}
{New physics scenarios\sep Collider physics\sep Extended scalar sectors \sep Dark Matter \sep Matter anti-matter asymmetry
}
\end{keywords}

\begin{figure}[h]
\begin{center}
	\includegraphics[width=9cm,height=6cm]{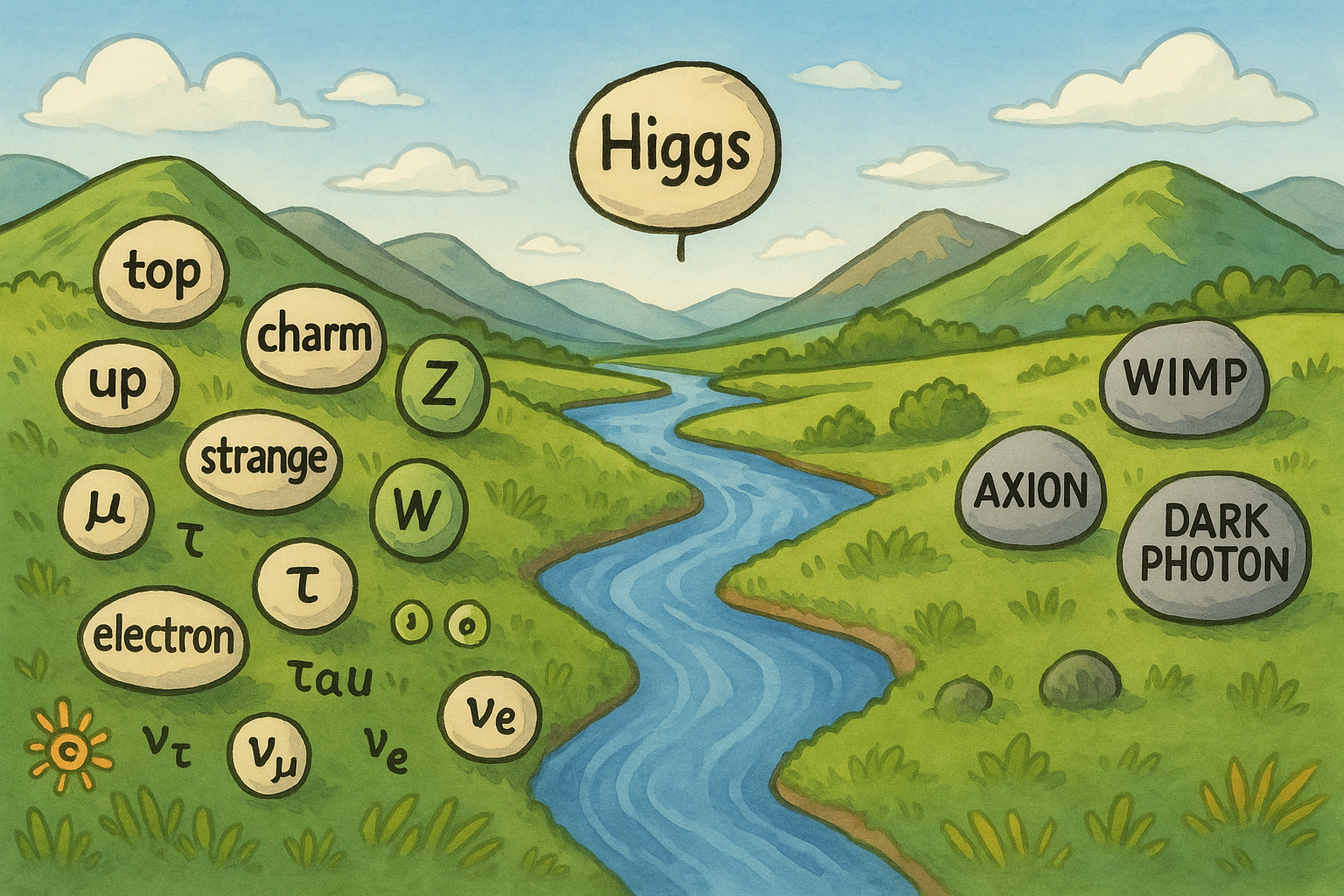}
	\caption{A ChatGPT vision of the portal between the visible world and the dark world.}
	\label{fig:titlepage}
    \end{center}
\end{figure}


\section*{Objectives}
\begin{itemize}
    \item Extended scalar sectors offer an appealing option to extend the scalar sector of the SM by introducing new particles. The corresponding matter content can help to explain several puzzles of the SM and can be tested at current or future collider facilities.
    \item Care must be taken when exploring the parameter space of such new scenarios, as it is subject to a large number of theoretical and experimental constraints.
    \item There are multiple options to extend the scalar sector of the SM, typically characterized by the transformation behavior under the SM electroweak gauge group. More parameters typically offer a larger choice of features and additional signatures.
    \item Models with extended scalar sectors are in particular good scenarios to introduce dark matter candidates, CP violation, or both. In turn, they can also allow for a strong first-order electroweak phase transition in certain cases.
\item A straightforward way to test these is the investigation at current of future collider facilities. Some regions of parameter space are already explored, but there is still enough room for investigation at such machines.
\end{itemize}

\section{Introduction}\label{intro}
{
The Standard Model (SM) of particle physics is build with a minimal scalar sector and a single spin zero complex field, an $SU(2)$ doublet  with four components, three of which will give mass to gauge bosons, after electroweak symmetry breaking (EWSB)~\cite{Englert:1964et,Higgs:1966ev,Guralnik:1964eu}. The Higgs field is a doublet of the group $SU(2)$ because this is the simplest object that accomplishes EWSB. This minimal set-up predicts only one Higgs boson that is consistent with the spin zero particle discovered by the ATLAS~\cite{ATLAS:2012yve} and CMS~\cite{CMS:2012qbp}  collaborations at the Large Hadron Collider (LHC) in 2012. 

Since the Higgs discovery, particle physics entered a new and exciting era of research. The first question one has to ask is whether the particle discovered by the LHC experiments is indeed the SM Higgs, or whether it is a particle from an enlarged new physics sector, an extension that would include additional fields and interactions. 
There are many options to extend the SM by additional particle content. In this chapter, we concentrate on models with added spin-0 fields, that transform as scalars under Lorentz transformations. These fields, however, can have different transformation properties under the $SU(2)\,\otimes\,U(1)$ gauge group of the SM, and are typically labeled according to their respective transformation properties. Depending on these properties, the additional particle content can be neutral or carry electromagnetic charges or any other new charge related to some new conserved current. The new term in the Lagrangian, that we will just refer to as the potential, may or not conserve charge conjugation (C), parity (P) or their product which we call CP. In that case the particles may or not be C, P or CP eigenstates at the level of the potential. One should note however that since the SM does not conserve CP, these states will be at most approximate CP-eigenstates.

When building the potential we can always choose it to be P-conserving and therefore only C can be broken either explicitly, by terms in the Lagrangian, or spontaneously by a C-violating vacuum expectation value (VEV)~\cite{Cvetic:1992wa, Haber:2022gsn}. In this case the potential will also be CP-violating. The addition of new sources of CP-violation is one of the reasons for extending the Higgs sector as it provides new sources of CP-violation. The potential with the new fields can also be invariant under a new transformation. This in turn can lead to stable particles that can serve as dark matter candidates.

Once the quantum numbers of the new scalar fields are defined, the couplings to the SM gauge bosons are automatically determined. In order to complete the model one needs to write the new piece of the Yukawa Lagrangian. While in the SM the absence of tree-level flavor changing neutral currents (FCNC) happens naturally, once more fields are added this is no longer true. A well-known case is the 2-Higgs doublet model where one more complex doublet is added to the SM field content~\cite{Lee:1973iz}. In this case, if one writes the most general version of the model, tree-level FCNCs will be present such as for instance vertices involving a top and charm quark and a scalar. These FCNC currents are very constrained by experiment~\cite{Gersabeck:2012xk} and should therefore be avoided or at least constrained. This means that parameters of the Lagrangian that are related to tree-level FCNC currents are either zero, because they are protected by a symmetry, or small enough to be in agreement with experiment. The simplest way to remove tree-level FCNCs from the theory is by imposing some ad-hoc symmetry under which the Yukawa Lagrangian is invariant. A $\mathbb{Z}_2$ symmetry was used in the reference works~\cite{Glashow:1976nt, Paschos:1976ay}.

Any extension of the SM is a quantum field theory and therefore infinities will arise in calculations beyond Leading Order (LO). In this chapter we will focus only on theories with operators with dimension less or equal to four. This means that there are no coupling constants with inverse energy dimension. Since all terms compatible with the symmetries are included~\footnote{Soft breaking terms, that is, term of dimension smaller than four that to not respect the symmetries of the model, can be included as they do not spoil its renormalizability.} the calculations of physical observables in extended sectors are most of the times performed at leading order. Because the models are just being tested, precision is usually not an issue~\footnote{This is not true for some processes like for instance Higgs production via gluon fusion where it is known that QCD corrections are important for the production of any scalar. Similarly, corrections in the electroweak sector can in some scenarios lead to large enhancements of scalar self-couplings.}. However, contributions to physical observables measured with great precision, like SM-like Higgs observables, could have contributions from particles and interactions from an extended scalar sector. In such cases the new theory has to be properly renormalized (for a review on electroweak radiative corrections see \cite{Denner:2019vbn}).

\subsection{Open questions of the Standard Model}

The SM has provided a solid framework for interpreting the world we live in at the quantum level. Experimental data is in quite good agreement with the predictions of the SM, as testified by a number of experimental results including the current measurements performed by the LHC experiments (see e.g. \cite{atlgenpub,cmsgenpub}). In fact, except for some discrepancies that tend to disappear as more data is gathered and analyzed, all measured observables are in very good agreement with the SM predictions. Moreover, searches for new particles in many different extensions of the SM have not provided a single solid hint of physics beyond the SM.

There are however some observations and features that cannot be explained by the SM and for which an extension of its scalar sector could be a way out. 
The most prominent example is the existence of dark matter (DM), which cannot be explained with the particle content of the SM alone. If DM is indeed a particle, the SM has no suitable candidate compatible with all observations. The matter-antimatter asymmetry of the universe is another outstanding problem one needs to address. 
When extending the scalar sector of the SM, one has in mind the goal of providing at least one viable DM candidate and new sources of CP-violation that can explain baryogenesis~\cite{Sakharov:1967dj} . Searches for CP-violation and DM are among the top priorities of the future LHC runs. If in turn these problems are indeed solved by an enlarged Higgs potential, its shape will most certainly be probed in final states with two or three Higgs bosons at the LHC or any future colliders. Extended Higgs sectors are also constructed to handle several alternative ways of generating neutrino masses.

Usually, new physics models are called to rescue at least some of these open points. In the following, we will discuss several extensions of the scalar sector that allow to at least address some of the open questions. As previously stated we concentrate on models that are renormalizable, in the sense that couplings have mass dimensions of $\geq\,0$ in natural units. This naturally limits the number of possible terms in the potential. Imposing additional symmetries typically further reduces the number of terms and thereby additional parameters.
Any new extended theory build to tackle these questions also needs to be in agreement with all experimental results in particle physics.
These issues will be discussed in the next section.

\section{Constraints on extended scalar scenarios}\label{sec:constraints}

 As previously stated, the models being discussed are renormalizable and invariant under a given set of symmetries which include as a minimal set the SM gauge group $SU(3) \times SU(2) \times U(1)$. All new physics scenarios are subject to a number of theoretical and experimental constraints. We will present a short overview of the most relevant constraints imposed when discussing an extension of scalar sectors of the SM. It is common to split the constraints into theoretically motivated, that need to be obeyed from a systematic  viewpoint, and experimental constraints, where the latter require comparison with current experimental findings and should therefore be modified if new data becomes available. For simplicity we consider only the lowest order in perturbation theory and omit a discussion of higher-order corrections for both theoretical bounds or predictions that call for comparisons with experimental constraints if not mentioned otherwise.

\subsubsection*{Theoretical constraints}

Theoretical constraints are not all equally relevant. Since the models are built in a quantum field theory framework, the Higgs potential has to be bounded from below, which is equivalent to say that the vacuum has to be stable. If this was not the case, the vacuum could be unstable and decay into a state of lower energy. In somehow simpler terms the potential energy of a physical system is not allowed to become infinitely negative.   
This condition imposes what are called positivity conditions on the parameters of the potential. Another important condition is that the vacuum should not break the remaining $U(1)$ symmetry corresponding to electric charge. In the SM this is trivially accomplished because there is only one Higgs doublet. In models with only singlet extensions, again the symmetry cannot be broken if the singlets transform trivially under the SM gauge groups. However, as soon as doublet are added the parameter space will be in principle curtailed. As shown in~\cite{Ferreira:2004yd, Barroso:2005sm, Maniatis:2006fs, Ivanov:2007de} when only one more doublet is added if the universe is in the non-charge breaking minimum, this is the absolute minimum and any other stationary points above it are saddle points. As we add more doublets and/ or singlets, conditions of boundedness from below may become notoriously difficult to find. In fact, analytic conditions are known only for very simple cases. A discussion of the vacuum condition of models with N Higgs doublets can be found in~\cite{Ivanov:2010wz, Ivanov:2010ww}. 

Unitarity is another fundamental constraint as it forces the sum of probabilities in a scattering process to be equal to one. This in turn imposes bounds on masses and couplings of the models. One can test the unitarity of the model making use of perturbative unitarity. In this approach, all scalar $2\,\rightarrow\,2$ scattering processes are considered at leading order. The scattering amplitudes are then rewritten using a partial wave decomposition. This renders an $n\,\times\,n$ matrix of lowest order partial wave amplitude coefficients. One can then impose a constraint on the largest eigenvalue of the matrix to ensure unitarity.   

Finally, as all calculations of cross sections and branching ratios in QFT are perturbative, one needs to make sure that the expansion on a given set of couplings do not diverge. This forces some perturbativity bounds on the parameters of the potential at a given scale. Renormalization group equations (RGEs) can also be used to understand how the model parameters evolve with energy. This in turn could uncover inconsistencies of the new theory at higher energies. RGEs can be used to examine electroweak vacuum stability or the perturbativity of couplings at higher scales.

\subsubsection*{Experimental constraints}

New physics scenarios are equally subject to a large number of experimental constraints. This involves the calculation of these quantities and the extraction from a given experiment, and the subsequent comparison of the calculated values with current experimental data. Here, both theoretical predictions can change, e.g. when updating input parameters or including higher-order contributions, as well as experimental data can change when e.g. new measurements become available or experimental uncertainties are changing.

Depending on the particle content of the new physics model, several observables need to be investigated. Presently we consider the most relevant to be:
\begin{itemize}
\item{} {\bf Electroweak precision observables (EWPOs) }: 
These are a set of high accuracy measurements of electroweak interactions, mainly from $Z$ and $W$ boson observables to test the SM and BSM physics via quantum effects. In practice, this is done with a comparison between electroweak precision observables and the so-called oblique parameters $S,\,T,\,U$ \cite{Altarelli:1990zd,Peskin:1990zt,Peskin:1991sw,Maksymyk:1993zm}. These variables were derived in order to obtain a simple means to parametrize new physics contributions in the electroweak sector via modifications to the electroweak gauge boson propagators. The current values are made available by fitting collaborations such as e.g. GFitter\cite{Flacher:2008zq} (see also \cite{Haller:2022eyb}) or as obtained by the Particle Data Group \cite{ParticleDataGroup:2024cfk} using the code presented in \cite{Erler:1999ug}\footnote{The authors thank J. Erler for useful discussions regarding this point.}. Alternatively, one can compare to one specific observable that has been very precisely measured, emphasizing the precision of said variable to have an extremely strong impact. An example for such variables are e.g. the $W$-boson mass or the anomalous magnetic momentum of the muon.
\item{}{\bf Particle widths}: Since new models will contribute via quantum corrections to SM particle widths, one should check agreement with the experimentally measured width, once the new particle contributions are included. This is especially important for the electroweak bosons with masses and widths measured with great precision~\cite{ParticleDataGroup:2024cfk}. This constraint can however be used for all SM particles.  
Equally important, especially for models where particles with masses below half the Higgs mass are allowed, is the width of the 125 \GeV~ resonance. Although not known to very high precision, it renders a valuable constraint from possible on-shell decays into additional light scalar particles. The total decay width can also be related to the branching ratio of Higgs to invisible, a quantity that is currently mainly constrained from measurements by the LHC experiments.
\item{}{\bf Flavor observables}: Models that contain new scalars carrying electromagnetic charges can experience important constraints from flavor observables, as e.g. $B\,\rightarrow\,X_s\,\gamma,\,\Delta M_s$, or $B_s\,\rightarrow\,\mu^+\,\mu^-$, where the additional particles contribute at either leading order or through quantum corrections.  Recent averages can e.g. be obtained from the Heavy Flavour Averaging Group \cite{hflav}.

\item{}{\bf Collider measurements and searches}: Another important topic is the agreement of any new physics scenario with current measurements and searches performed at colliders and in particular at the LHC. A new physics scenario has to entail one neutral scalar with a mass $\sim\,125\,\GeV$ and with a large CP-even component. Furthermore, direct searches for scalar resonances decaying into various scalar states also exclude specific regions in the parameter space of any specific extended scalar sector. We will come back to this subject in section~\ref{sec:current}. Note that for many scalar models, in particular the ones with just added singlets and doublets under the SM gauge group, there are sum rules \cite{Gunion:1990kf} that link the couplings of the scalar particles to gauge bosons in the model according to
\begin{\eqn}\label{eq:sumrule}
\sum_i g_{h_i}^2\,=\,g_{\text{SM}}^2,
\end{\eqn}
where the sum goes over all neutral scalar particles in the theory and $g_{h_i}\,\lb g_{\text{SM}}\rb$ are the couplings of scalar $h_i\,\lb h_{125} \rb$ to electroweak vector bosons, respectively.
Therefore, many models with extended scalar sectors are already severely constrained by the current measurement of the 125 \GeV~ signal strength. Note that this sum rule is modified in the presence of doubly charged scalars, such that larger single couplings to vector bosons are allowed.
\item{}{\bf Dark matter related observables}: If the new physics scenario contains a DM candidate, physical observables have to be in agreement with DM experiments. The most relevant example is the DM relic density measurement, 
that in order avoid overclosure of the universe must not exceed the observed value $\Lambda_{CDM}$ h$^2 \leq 0.120$, according to the Planck collaboration \cite{Planck:2018vyg}. Overclosure could be caused by the excessive gravitational pull. The other relevant constraints come from direct and indirect detection measurements. All DM constraints are discussed in detail in chapter~\ref{DM}.

\item{}{\bf CP-violating observables}: One of the most important reasons to add a new scalar sector to the theory is to increase the amount of CP-violation of the SM in order to explain baryon asymmetry. There are extremely precise low energy experiments such as the experimental measurement of the Electric Dipole Moment of the electron~\cite{Roussy:2022cmp} and of the neutron~\cite{Abel:2020pzs}. When the new sector is CP-violating new contributions to these observables arise, especially from the new CP-violating Yukawa couplings, leading to severe constraints on the models.
There are also direct constraints coming from the measurements of the ratios between the CP-even and the CP-odd components of the Yukawa couplings. 
So far ATLAS and CMS have performed a direct measurement of the ratio of the CP-odd and CP-even components of the Yukawa couplings for the top Yukawa~\cite{ATLAS:2020ior} (in the production) and for the tau~\cite{CMS:2021sdq} (in the decays). Probing other Yukawa couplings is
one of the goals of future LHC searches.

\end{itemize}

\noindent
Various tools are on the market that allow for relatively easy comparisons of new physics scenarios, several focusing on scalar extensions, with the above-mentioned constraints. We will comment on these in section~\ref{sec:tools}.

\section{Singlet extensions}\label{sec:singlets}

In singlet extensions, the SM scalar sector is enhanced by one or more real or complex fields that transform as singlets under the SM gauge group. Due to the trivial transformation properties, the kinematic terms for a single scalar field $\phi$ are simply given by
\begin{\eqn*}
\mathcal{L}\,\supset\lb \partial_\mu \phi \rb \lb \partial^\mu \phi \rb^\dagger
\end{\eqn*}
in such scenarios.

In general, there are several variations for these extension terms. In order to preserve renormalizability, for an extension with $n$ scalar fields $\phi_i$, only terms of the form 

\begin{equation}
    V(\phi_i, \Phi) =V_{\text{singlets}} (\phi_i,\Phi) + V_\text{SM}(\Phi),
    \label{eq:generalpotential}
\end{equation}
where
\begin{equation}
    \begin{aligned}
        V_{\text{singlets}} (\phi_i, \Phi) & =  a_i \phi_i
        + m_{ij} \phi_i \phi_j
        + T_{ijk} \phi_i \phi_j \phi_k
        + \lambda_{ijkl} \phi_i \phi_j \phi_k \phi_l                              
                                              + T_{iHH}\phi_i (\Phi^\dagger\Phi)
        + \lambda_{ijHH}\phi_i\phi_j(\Phi^\dagger\Phi)                                  \\
    \end{aligned}
\end{equation}
are allowed, {in the case of real singlets}. In the above expression, $\Phi$ denotes the SM-like doublet that is responsible for electroweak symmetry breaking.

The number of free parameters in the above expression can, as discussed, be further reduced by introducing additional symmetries. In case e.g. an additional adequate $\mathbb{Z}_2$ symmetry is introduced, and if the respective scalar field does not break the symmetry by developing a non-zero vacuum expecation value, the respective mass eigenstate cannot decay into SM particles and is therefore stable, being a possible candidate for dark matter. Whether such a scenario is able to explain the DM content of the universe depends on the specific setup and parameter choices of the model.

As a simple} example, we consider a scenario where a real singlet is added to the SM scalar sector, which corresponds to the above expression with $n=1$. In addition, in order to further reduce the number of free parameters, one can also introduce said $\mathbb{Z}_2$ symmetry ($\phi_1\,\rightarrow\,-\phi_1$) for the new scalar, where all other fields are even under this transformation. We furthermore assume both fields to obtain a vacuum expectation value.

Independently of the number of free parameters in the potential, the resulting phenomenology is determined by the following free parameters:
\begin{\eqn}\label{eq:singpars}
m_H,\,\sin\al,\,\Gamma_H,
\end{\eqn}
where for simplicity we have now assumed that the second scalar is heavier than the 125 \GeV resonance (the inverse case follows analogously). Here, $m_H$ and $\Gamma_H$ are the mass and the total width of the heavy scalar, while $\sin\al$ is the mixing angle that transforms from gauge to mass eigenstates. In the scenario discussed above,
\begin{\eqn*}
\Gamma_H\lb m_H\rb\,=\,\lb \sin^2\,\al\rb \,\Gamma_H^\text{SM}\,\lb m_H\rb\,+\,\Gamma_{H\,\rightarrow\,h\,h}
\end{\eqn*}
where $\Gamma_H^\text{SM}\lb m_H\rb$ is the total width of a SM-like scalar at mass $m_H$. Note that the  parameters in eqn. (\ref{eq:singpars}) can be easily exchanged by other inputs, see e.g. discussion in \cite{Robens:2016xkb}.

An important view of the parameter space can e.g. be given by considering the $\lb m_H,\,\sin\al\rb$ plane. The mixing angle $\sin\al$ is of special importance, as all couplings to SM-like particles of the new scalar (apart from couplings in the scalar sector) are rescaled by $\sin\al$ with respect to their SM-like value according to $g_H\,=\,\sin\al\,g_H^\text{SM}$, where again $g_H^\text{SM}$ corresponds to the SM-like coupling for the scalar $H$. Similarly, the couplings of the lighter scalar are suppressed by $\cos\al$. 

\begin{figure}[h!]
        \begin{center}
            \includegraphics[width=0.45\textwidth]{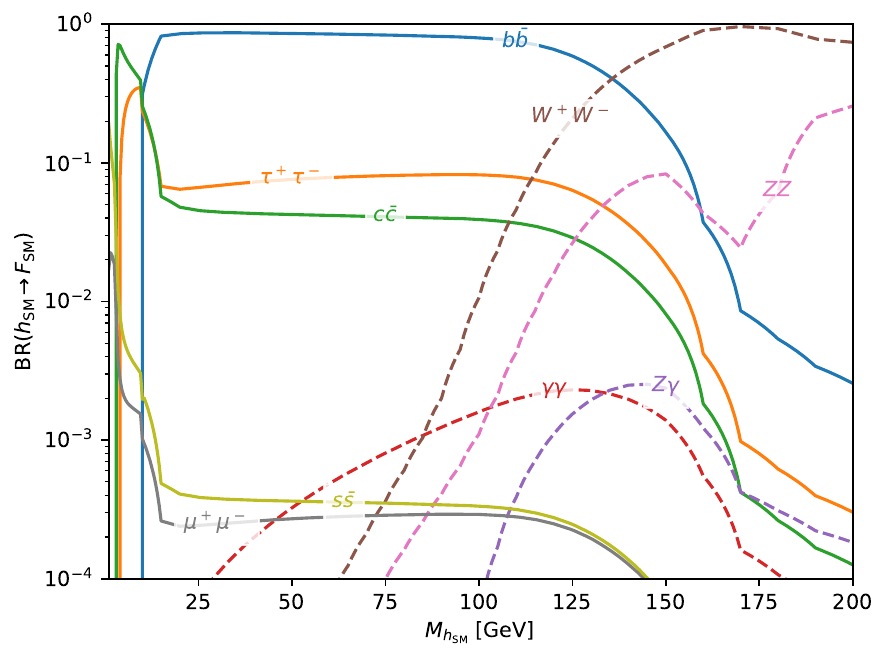}
             \includegraphics[width=0.45\textwidth]{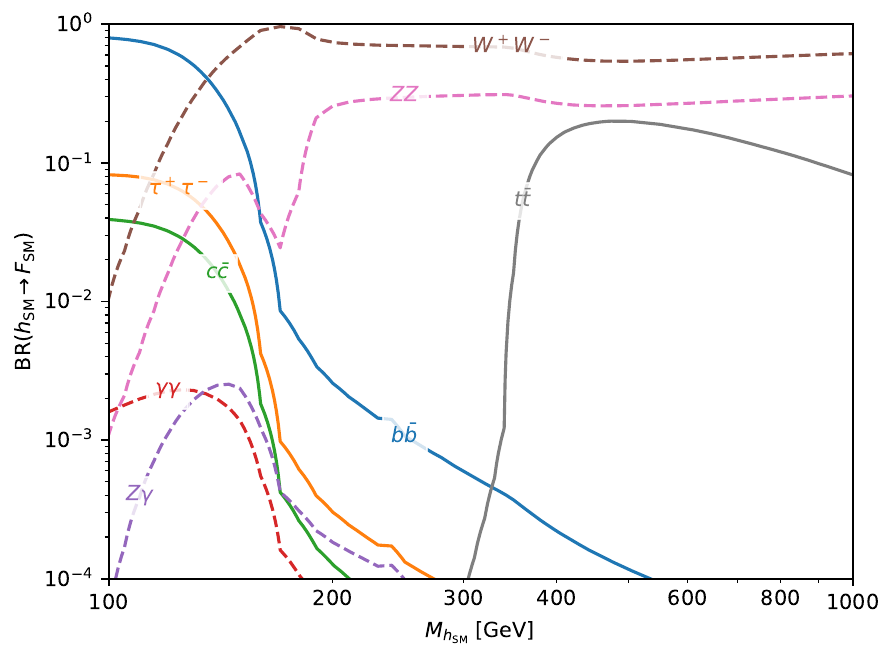}
            \caption{\label{fig:brs} Branching ratios of an SM-like scalar assuming no new physics decay channel, depending on the scalar mass, taken from \cite{Robens:2019kga}. {\sl (Left)} for masses $\leq\,200\,\GeV$; {\sl (right)} for masses $\geq\,100\,\GeV$. }
        \end{center}
        
    \end{figure}

Considering the sum rule  given by eqn.~ (\ref{eq:sumrule}), we see that from the signal strength measurements alone we already get relatively strong constraints on $|\sin\al|$. Furthermore, direct searches for heavy resonances additionally can constrain the allowed regions. Here, decay rates for the new scalars typically follow the decays of an SM-like particle of the same mass, unless new physics channels (e.g. $H\,\rightarrow\,h\,h$) open up. Branching ratios for the latter decay can typically be tuned by specific choices of the new physics parameters.

In figure \ref{fig:brs}, we show the branching ratios of SM-like scalars for various mass ranges. The corresponding predictions were obtained from \cite{LHCHiggsCrossSectionWorkingGroup:2016ypw} and produced using HDecay \cite{Djouadi:1997yw,Djouadi:2018xqq}.

It is extremely important to understand that in experimental searches for new physics scenarios, the total decay widths for the new scalars are not arbitrary parameters but can have an important impact on the interpretation of new physics searches. For simple singlet extensions, this has been know for quite a while. We refer the reader e.g. to \cite{Kauer:2015hia,Kauer:2015dma,Dawson:2015haa,Carena:2018vpt,Kauer:2019qei,DiMicco:2019ngk,Banfi:2023zav,Feuerstake:2024uxs} for various recent discussions. Modification of the width by hand, e.g. away from the model predictions, can lead to arbitrarily large or small branching ratios. Similarly, the inclusion of finite width effects can have an important impact on the differential distributions that are realized within the different model scenarios.

\begin{center}
\begin{figure}
\begin{center}
\includegraphics[width=0.45\textwidth]{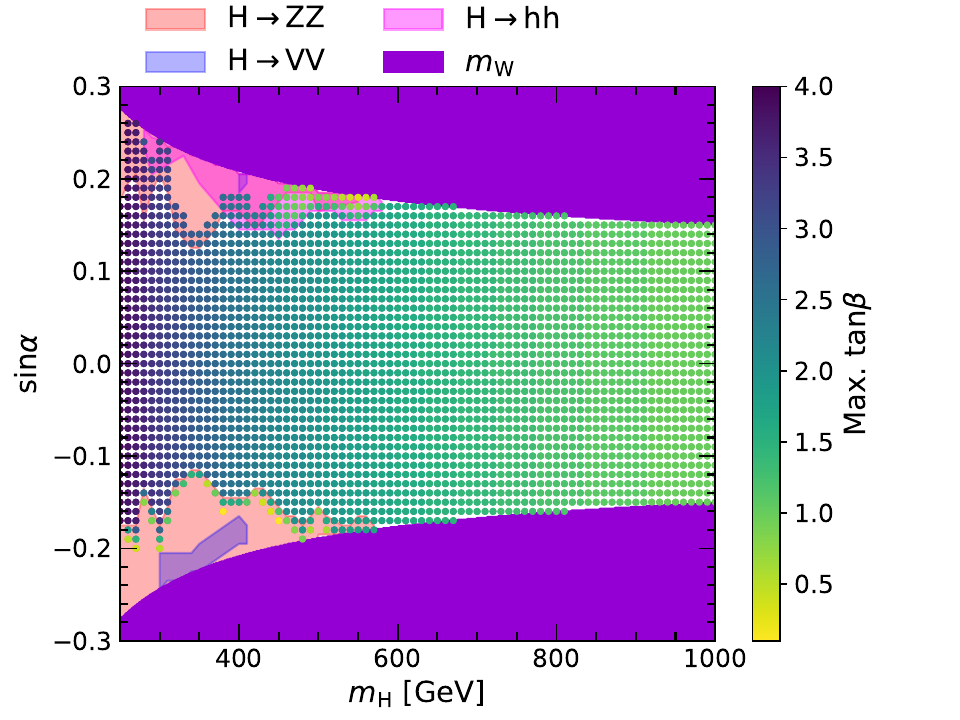}
\includegraphics[width=0.45\textwidth]{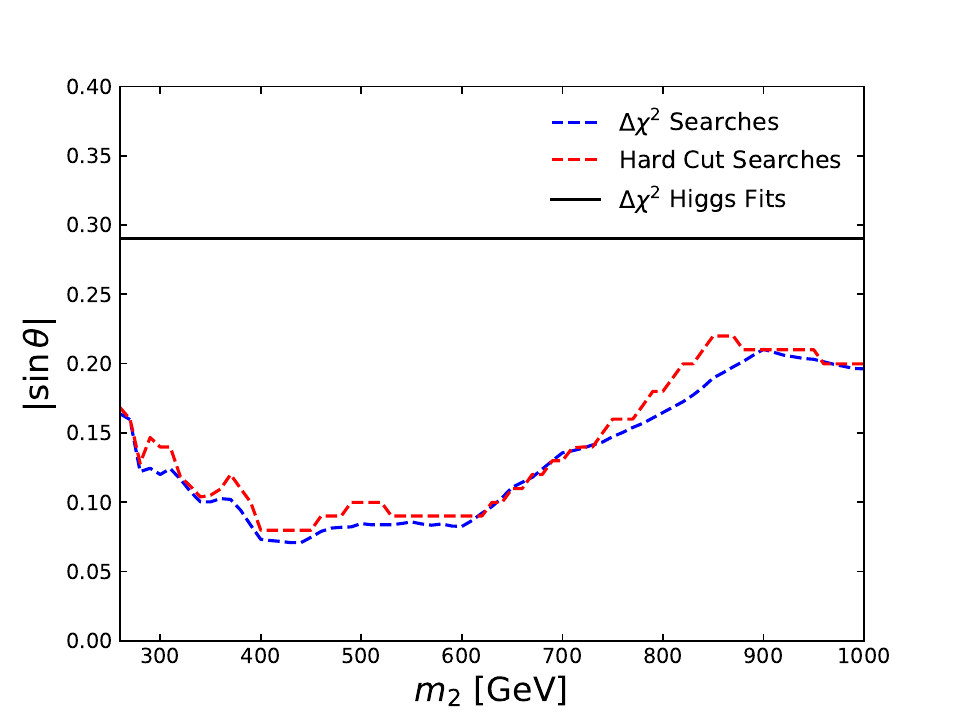}
\caption{\label{fig:sings} Allowed and excluded regions in the $\lb m_H,\,\sin\al\rb$ plane with {\sl (left)} and without {\sl (right)} a $\mathbb{Z}_2$ symmetry (in this case the mixing angle is denoted by $\sin\theta$), taken from \cite{Feuerstake:2024uxs} and \cite{Lewis:2024yvj}, respectively. For the left plot, constraints were obtained using the code developed in \cite{Robens:2015gla}, with an updated interface to HiggsTools \cite{Bahl:2022igd}. See \cite{Lewis:2024yvj} for details on the scan for the scenario without a $\mathbb{Z}_2$ symmetry. In particular, the $\Delta \chi^2$ and ``hard cut'' are two methods to incorporate limits form searches for resonant scalar searches \cite{Adhikari:2020vqo,Lane:2024vur}, with ``hard cut'' being the traditional method.}
\end{center}
\end{figure}
\end{center}

In figure \ref{fig:sings}, we display the allowed parameter space in the real singlet extension with and without a $\mathbb{Z}_2$ symmetry.

Of course, there is in principle no limit to add more (complex) singlet fields to the SM Lagrangian. A prime example is the case of a complex singlet, which was first discussed in \cite{Barger:2008jx}; 
see e.g \cite{Ferreira:2016tcu,Chiang:2017nmu,Cheng:2018ajh,McDowall:2018tdg,Grzadkowski:2018nbc,Chen:2019ebq,Chiang:2019oms,Ghosh:2021qbo,Cho:2021itv,Cho:2023hek,Zhang:2023mnu,Lane:2024vur} for more recent discussions, also including options for dark matter scenarios to be discussed in section \ref{DM}.

It should be noted that without any additional symmetries imposed, a model with one complex singlet and two real singlets are equivalent, see e.g. the discussion in the appendix of \cite{Robens:2019kga}. Models with one complex or two real additional singlets allow for interesting novel production and decay modes of the form 
\begin{\eqn}\label{eq:novsig}
p\,p\,\rightarrow\,h_i\,\rightarrow\,h_j\,h_j,\; p\,p\,\rightarrow\,h_i\,\rightarrow\,h_j\,h_k
\end{\eqn}
where in the first process none of the scalars corresponds to the 125 \GeV~ resonance. These channels have first been proposed and discussed in \cite{Robens:2019kga} in the context of a model with two additional $\mathbb{Z}_2$ symmetries; recent updates on the available parameter planes are given e.g. in \cite{Robens:2022nnw,Robens:2025tew}. First searches of the LHC experiments for such final states can be found in \cite{CMS:2022qww,CMS:2022suh,ATLAS:2023tkl,CMS:2023boe,CMS:2024bds,CMS:2024phk,ATLAS:2024auw,ATLAS:2024itc,CMS:2025aql,CMS:2025anz,ATLAS:2025rfm}.  Note that processes of the form given in eqn. (\ref{eq:novsig}) can appear in any model that contains 3 CP even neutral scalar bosons, as e.g. also realized in next-to-minimal supersymmetric extensions of the SM.

\section{Two Higgs Doublet Models}

\subsection{The CP-conserving softly broken $\mathbb{Z}_2$ symmetric 2HDM}

Two Higgs Doublet Models (2HDMs) are one of the most popular extensions of the SM scalar sector. An initial motivation was that when going to supersymmetric scenarios, a second scalar doublet is needed in order to preserve supersymmetry and at the same time give masses to all fermions. In 2HDMs the scalar sector is enhanced by a second complex scalar field that acts as a doublet under the SM gauge group. A typical notation is then given by
\begin{\eqn*}
\Phi_a\,=\,\lb \begin{array}{c}{\phi^+_a}\\{\lb v_a+ \phi_a\,+\,\imath\,\eta_a \rb/\sqrt{2}}  \end{array}\rb
\end{\eqn*}
where $a\,=\,1,2$, $v_a$ denotes the vacuum expectation value of the respective doublet, and the fields $\phi^+_a\,\lb \rho_a,\,\eta_a\rb$ are taken to be complex (real) respectively. After electroweak symmetry breaking, we are left with 5 physical scalar fields, denoted by
\begin{\eqn*}
\underbrace{h,\,H,}_{\text{neutral, CP even}} \underbrace{A}_{\text{neutral, CP odd}}, H^\pm,
\end{\eqn*}
where we already assumed that CP is conserved. The most generic renormalizable form of the model allow for both flavour changing neutral currents as well as CP violation. In order to forbid the former,  additional symmetries are imposed on the potential. We refer the reader to e.g. \cite{Branco:2011iw} and references therein for an overview on possible symmetry classes, as well as a general exhaustive overview on 2HDMs.

The most common version of the 2HDM, used as benchmark in many experimental searches, is the softly broken $\mathbb{Z}_2$ symmetric version of the model with the potential given by
\begin{align}\label{eq:2hdmpot}
    V_\text{2HDM} =&\ \mu_1^2 |\Phi_1|^2 + \mu_2^2 |\Phi_2|^2 - \mu_{12}^2\,\lb \Phi^\dagger_1\,\Phi_2+ \text{h.c.}\rb+ \frac{1}{2} \lambda_1 |\Phi_1|^4 + \frac{1}{2} \lambda_2 |\Phi_2|^4 + \lambda_3 |\Phi_1|^2 |\Phi_2|^2 + \lambda_4 |\Phi_1^\dagger \Phi_2|^2 \nonumber\\
    &+ \frac{1}{2} \lambda_5 \left[(\Phi_1^\dagger\Phi_2)^2+\text{h.c.}\right]\,. 
\end{align}
where all parameters are taken to be real. The potential has 8 free parameters which are usually chosen to be $v\,\equiv\,\sqrt{v_1^2+v_2^2}\,\sim\,246\,\GeV$,  know from electroweak precision measurements and either $h$ or $H$ has to mimic 
the 125 \GeV~ resonance measured by the LHC experiments. The remaining 6 free parameters are usually chosen to be
\begin{\eqn}\label{eq:2hdmpars}
m_{h/H},\,m_A,\,m_{H^\pm},\,\tan\,\be\,\equiv\,\frac{v_2}{v_1},\,\cos\lb \be-\al\rb,\,\mu_{12}^2
\end{\eqn}
where we introduced the mixing angles $\be$ and $\al$ that are related to mixings of the gauge and mass eigenstates in the CP-odd and even neutral sectors, respectively\footnote{Note this is basis dependent; we refer the interested reader e.g. to \cite{Davidson:2005cw} for more details.}.  The parameters in eqn.~(\ref{eq:2hdmpars}) are a typical parameter choice frequently used by the experimental collaborations. In particular, the mixing angle combination $\cos\lb \be-\al \rb$ is of particular relevance as the limits 0 and 1 correspond to the so-called alignment limit where one of the Higgs couples to electroweak gauge bosons as in the Standard Model. This is therefore the rescaling entering the sum rule given by eqn. (\ref{eq:sumrule}), and typically, as we will also show below, the value of this mixing angle is constrained to a region relatively close to 0 for $h$ and close to 1 for $H$.

While the softly broken $\mathbb{Z}_2$ realization of the 2HDM scalar potential is given by eqn. (\ref{eq:2hdmpot}), there is still some freedom in assigning couplings to fermions. There are four different Yukawa types that ensure flavour conservation. Type 1 denotes the scenario where all fermions couple to $\Phi_2$. In type 2, only up-type quarks couple to $\Phi_2$, while down-type quarks and lepton couple to $\Phi_1$. In lepton-specific scenarios, only leptons couple to $\Phi_1$, while quarks couple to $\Phi_2$. Finally, in flipped scenarios, only down-type quarks couple to $\Phi_2$, while up-type quarks and leptons couple to $\Phi_1$. 

\begin{figure}[htb!]
\begin{center}
\includegraphics[width=0.4\textwidth]{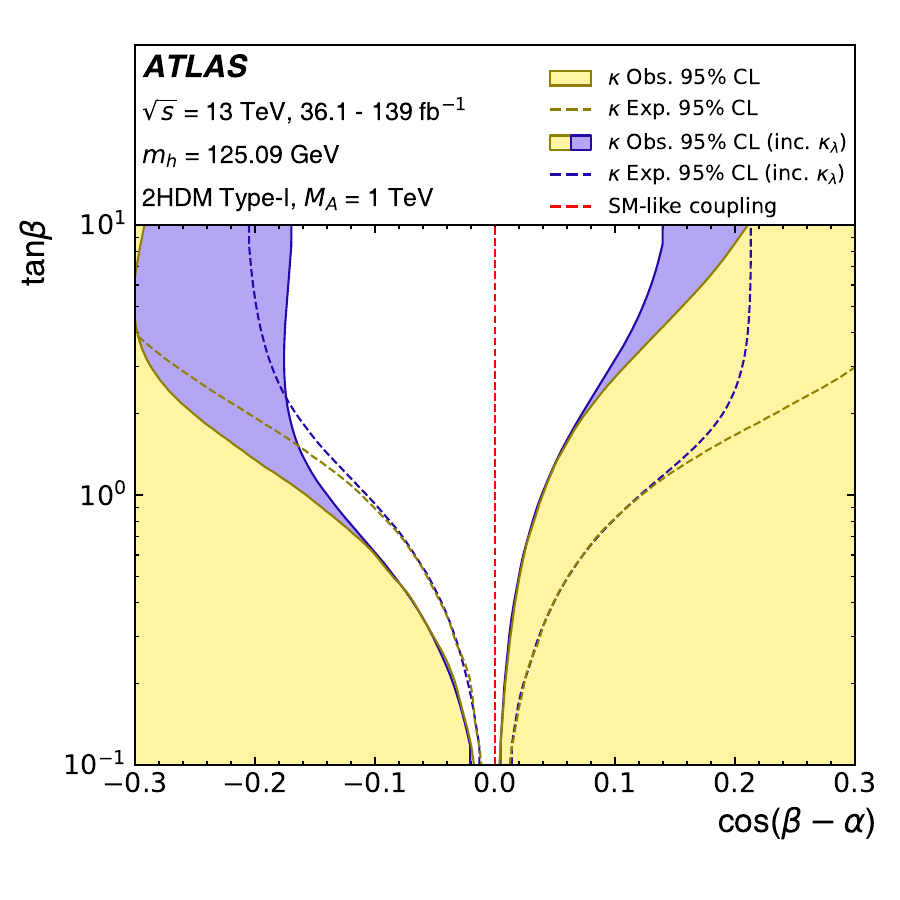}
\includegraphics[width=0.4\textwidth]{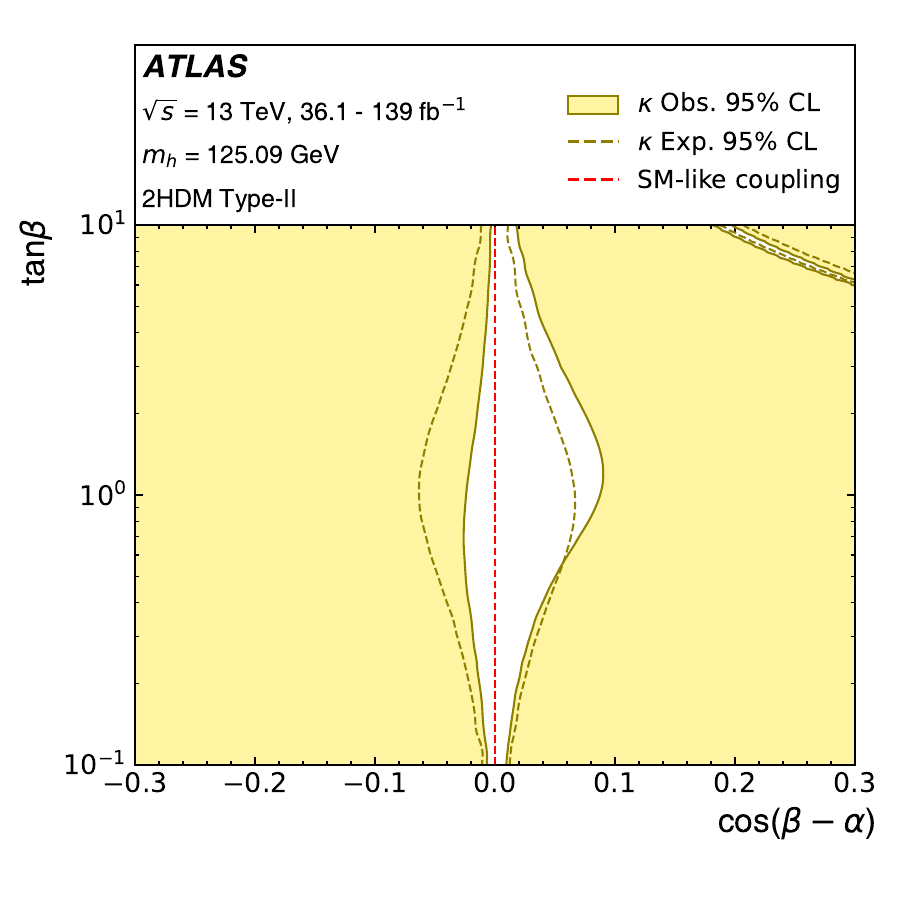}
\includegraphics[width=0.4\textwidth]{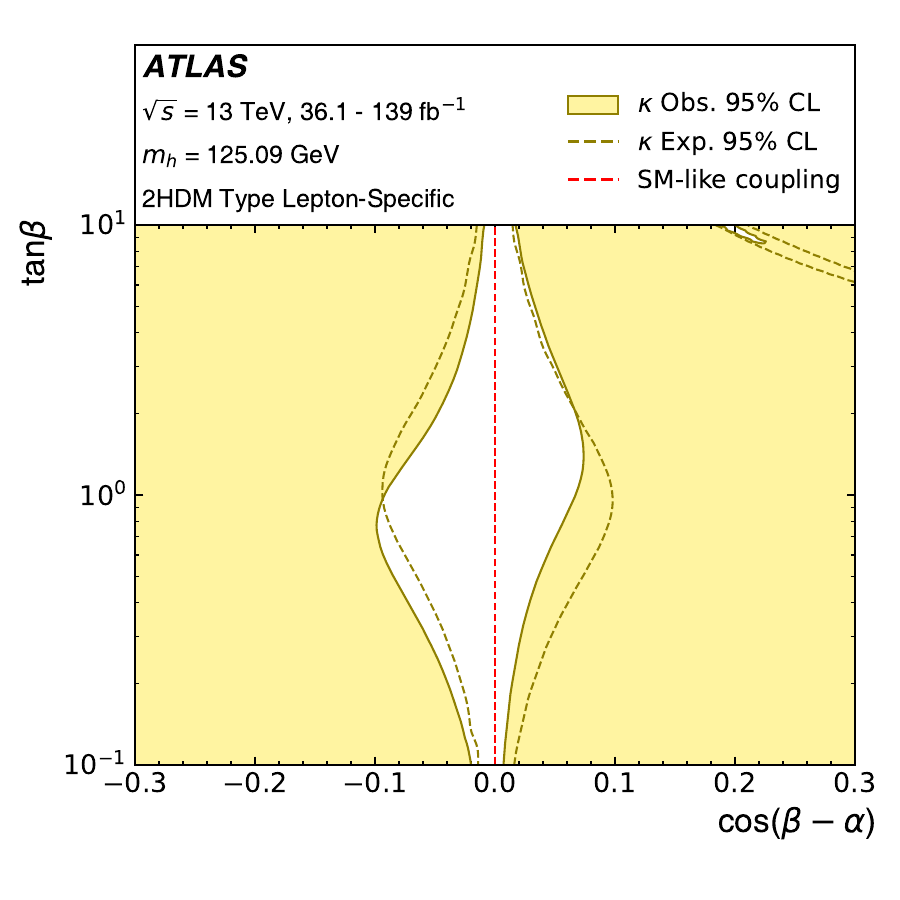}
\includegraphics[width=0.4\textwidth]{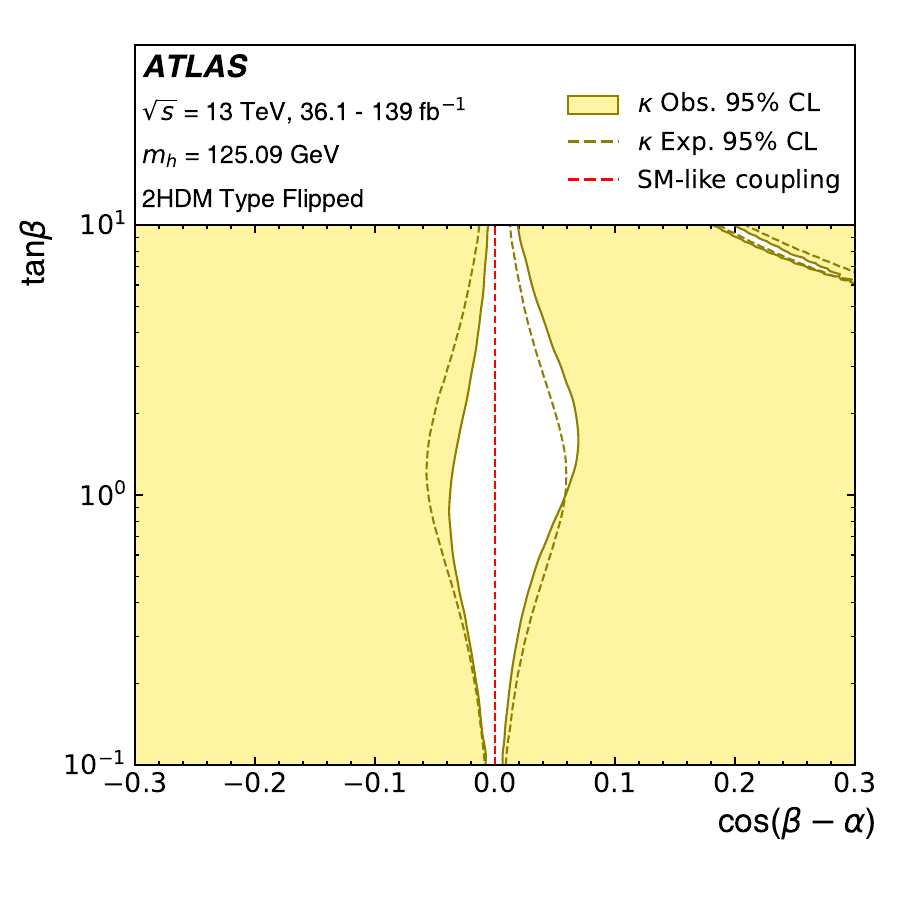}
\caption{\label{fig:atlcba} Signal strength fit constraints and allowed regions in the $\lb\cos\lb\beta-\alpha\rb,\,\tan\be\rb$ plane for various types of 2HDMs; ATLAS combination results. Taken from \cite{ATLAS:2024lyh}. 
}
\end{center}
\end{figure}

In general, it is not straightforward to find two-dimensional planes in a multi-dimensional parameter space that are clearly subject to a specific constraint. For 2HDMs, however, there are at least two examples where this is the case. One is the signal strength for the 125 \GeV~ resonance; measurements of this give clear constraints in the $\lb \cos\lb \be-\al \rb,\tan\be \rb$ plane. The most recent fits  from the ATLAS collaboration are shown in figure \ref{fig:atlcba}.

\begin{center}
\begin{figure}[htb!]
\begin{center}
\includegraphics[width=0.5\textwidth]{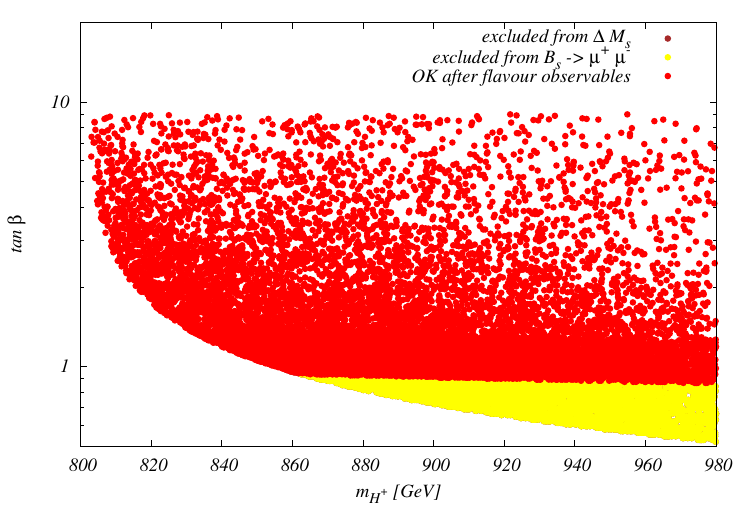}
\caption{\label{fig:brsg} Bounds on the $\lb m_{H^\pm},\,\tan\be\rb$ plane from B-physics observables, implemented via the SPheno \cite{Porod:2011nf}/ Sarah \cite{Staub:2013tta} interface,  and compared to experimental bounds \cite{CMS-PAS-BPH-20-003,HFLAV:2019otj}. The contour for low $\lb m_{H^\pm,\,\tan\be}\rb$ values stems from \cite{Misiak:2020vlo,mm}. Figure taken from \cite{Robens:2022jrs}.}
\end{center}
\end{figure}
\end{center}

Another prominent example are bounds in the $\lb m_{H^\pm},\tan\be \rb$ plane from flavor constraints. In figure \ref{fig:brsg} we show a plot derived in \cite{Robens:2022jrs} (update of a figure first presented in \cite{Robens:2021lov}) which exemplarily shows exclusion contours (current bounds might render slightly different constraints).
Other 2-dimensional constraints that are prominent can e.g. stem from electroweak precision observables that render clear constraints in 2-dimensional mass difference planes. One example for this can e.g. be also found in \cite{Robens:2021lov}. 

\subsection{The CP-violating softly broken $Z_2$ symmetric 2HDM}

The choice of complex $\lambda_5$ and $\mu_{12}^2$ leads to a potential that breaks CP explicitly. This is one of the simplest extension of the SM that breaks CP and is used by the LHC experiments as a benchmark model~\cite{Ginzburg:2002wt, Khater:2003wq, Biekotter:2024ykp}
The three neural states mix and are no longer be CP-eigenstates. They are usually defined as $h_1$, $h_2$ and $h_3$ and any of them can be the 125 GeV Higgs. This model has an additional very important constraint - the experimental value of the electron Electric Dipole Moment (EDM)~\cite{Roussy:2022cmp}. Although other related constraints can also play a role, like the neutron EDM, it is the electron EDM that curtails most of the parameter space of the CP-violating model. As discussed in~\cite{Cvetic:1992wa, Fontes:2015xva, Haber:2022gsn}, CP-violation in the scalar sector has its origin in C-violation, because it is always possible to choose the parity numbers in such a way that a gauge theory with scalars and gauge bosons is P-conserving. 

The important point now is how to search for CP-violation at a collider. If CP-violation has its origin in the scalar sector we would in principle need to find at least one new scalar. If three vertices of the type $h_1 h_2 Z$, $h_1 h_3 Z$ and $h_2 h_3 Z$ are present in the theory, this is a clear sign~\cite{Mendez:1991gp, Fontes:2015xva} of CP-violation at tree-level. This is true beyond the 2HDM. So these three physical processes among many others~\cite{Fontes:2015xva, Haber:2022gsn} can be used to probe CP-violation at colliders. Another possibility is to tie the same three vertices in a loop of a triple gauge boson interaction. This scenario will be discussed in more detail in the section devoted to DM and CP-violation. 
\begin{figure}[h!]
  \centering
  \begin{tabular}{cc}
    \includegraphics[width=0.45\textwidth]{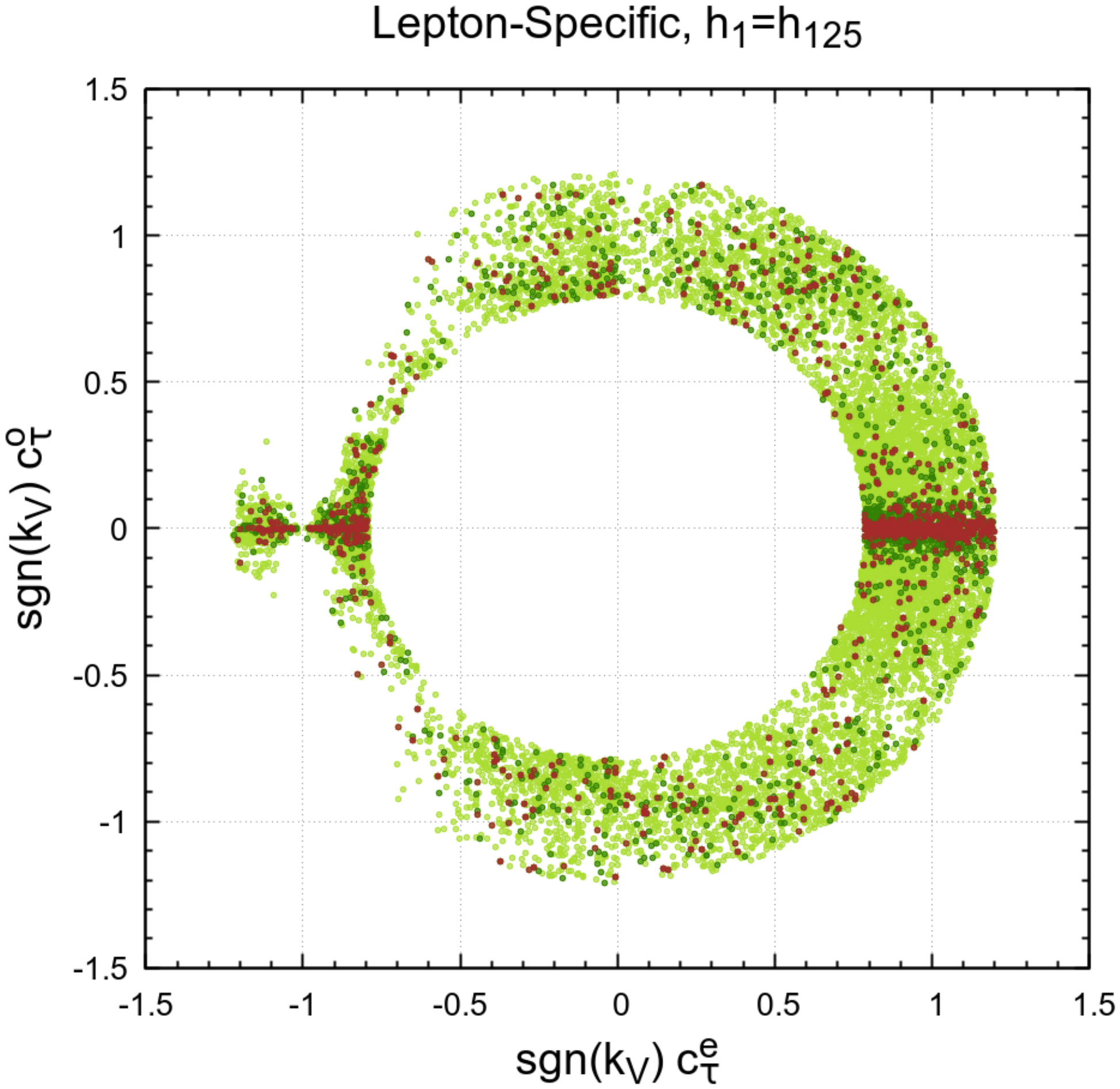}&
    \includegraphics[width=0.45\textwidth]{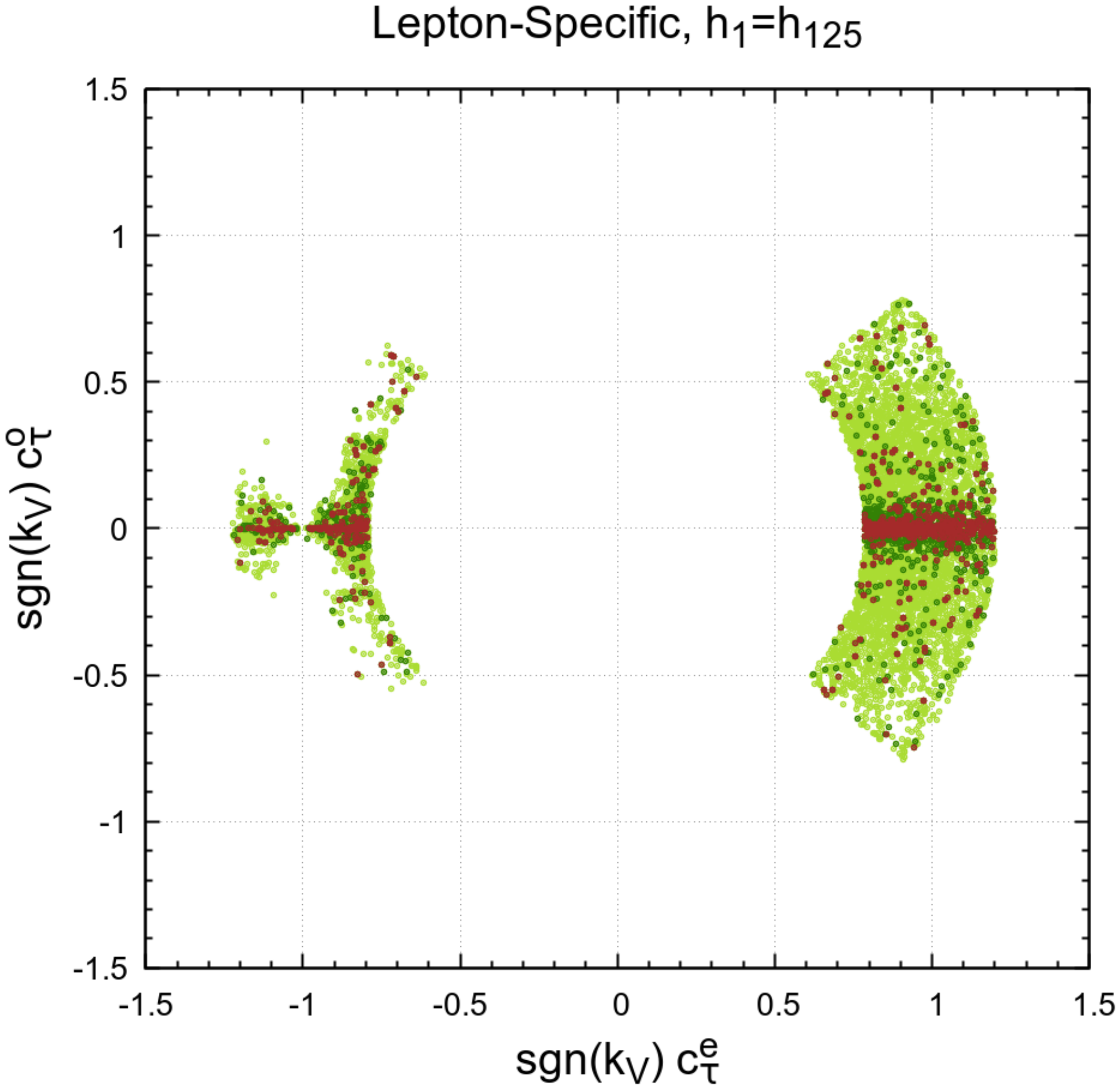}
  \end{tabular}
  \caption{Figure taken from~\cite{Biekotter:2024ykp}. CP-odd \textit{vs.}~CP-even component in the
    $h_{125}\tau\bar{\tau}$ coupling for the allowed
    parameter points in the LS
    model, assuming $h_1=h_{125}$,
    using 13~TeV LHC Higgs data on $h_{125}$ collected
    until 2022 and constraints from BSM
    scalar searches included in HT-1.1.3.
    In the left panel, the
    limit $\alpha_{h\tau\tau} < 41^\circ$
    from angular correlations of $\tau$ leptons
    in $h_{125} \to \tau \bar \tau$ decays
    is not applied, whereas the
    right panel includes this limit.  The light green points are consistent
    with the old eEDM of
    $8.7 \times 10^{-29} \,
    \text{e.cm}$~\cite{ACME:2013pal},
    the dark green points with the more recent ACME
    result $1.1 \times 10^{-29}
    \, \text{e.cm}$~\cite{ACME:2018yjb}.
    The dark red points obey
    the currently strongest limit on the
    eEDM $4.1\times 10^{-30} \,
    \text{e.cm}$ reported by JILA~\cite{Roussy:2022cmp}.
    The fermion masses in the loops
    of diagrams contributing to the eEDM
    were taken as pole masses.
    The limit $\alpha_{h\tau\tau} <
    41^\circ$~\cite{CMS:2021sdq} from
    searches for CP-violation in angular
    correlations of $\tau$ leptons in $h_{125} \to \tau
    \bar \tau$ decays has not been applied in either
    of the plots in this figure. sgn $(k_V)$ is the sign of the coupling modifier of the Higgs to the gauge bosons.} 
  \label{fig:3}
\end{figure}

When CP-violation occurs in the scalar sector the Yukawa couplings are also modified to the form $c_f^e + i c_f^o \gamma_5$, where $f$ is a generic fermion, the superscript $e$ stands for even while $o$ stands for odd. This current is C-even but the two terms have different P-numbers if parity is conserved.  Therefore this is a CP-violating Yukawa coupling. The coefficients $c_f^e$ and $c_f^o$ could in principle be all independent which means that one should probe all Yukawa couplings.    
In fact, some very interesting and unexpected scenarios can arise in these CP-violating extension of scalar sectors. There are limiting scenarios in these models with the possibility of having different ratios of pseudoscalar to scalar couplings in the Yukawas~\cite{Fontes:2015mea}. Depending on the Yukawa type it is possible to have a large pseudoscalar coupling to the leptons or down-type quarks while having a small pseudoscalar coupling to the up-quarks. This possibility shows the importance of measuring all Yukawa couplings independently. An analysis using all up-to-date experimental results was recently performed in~\cite{Biekotter:2024ykp}. So far ATLAS and CMS have performed a direct measurement of the ratio of the CP-odd and CP-even components of the Yukawa couplings for the top Yukawa~\cite{ATLAS:2020ior} (in the production) and for the tau~\cite{CMS:2021sdq} (in the decays).

We just present a particular example in figure~\ref{fig:3} for the Lepton-specific (LS) softly broken $\mathbb{Z}_2$ symmetric 2HDM. Since the LHC bounds are basically on total rates for given processes, these constraints set an upper and a lower bound on the function $(c_f^e)^2 + (c_f^o)^2 $, and therefore the allowed space is constrained between two circumferences. In some scenarios, and some Yukawa types, the constraints become so strong that part of the ring is dented or just disappears. This is even more so if the LHC constraints are combined with the electron EDMs experimental results. However, there are also direct measurements on the couplings. This is what can be seen on the right plot where the limit $\alpha_{h\tau\tau} < 41^\circ$~\cite{CMS:2021sdq} from angular correlations of $\tau$ leptons in $h_{125} \to \tau \bar \tau$ decays is applied.

\section{More general extensions}

{There is no limit on the number and group representation of the scalar fields to be added to the SM. Although the most studied extensions include mainly singlets, doublets and/ or triplets under the $SU(2) \otimes\, U(1)$ gauge group of the SM, all other representations are allowed and have been used in more exotic scenarios. As discussed above, the new physics scenario needs to provide at least 1 CP-even neutral scalar that complies with the properties of the 125 \GeV~ resonance discovered by the LHC experiments. Furthermore, it needs to provide a viable mechanism and a correct pattern of electroweak symmetry breaking, and to obey all constraints listed in section~\ref{sec:constraints}. The latter typically set some requirement on the additional free parameters of the scalar potential. In some scenarios, higher-order calculations are needed in order to bring the model predictions in agreement with current measurements~\footnote{A prime example for this are supersymmetric scenarios, which at leading order predict a Higgs mass $\leq\,m_Z$. Here, higher-order corrections are needed in order to level the mass to the currently measured value, see e.g. \cite{Martin:1997ns}.}.

As discussed above, models with additional field content can originate very different phenomenology from the one expected for the SM. One can easily produce models with dark matter, with CP-violation in the scalar sector, with charged or even doubly charged scalars, among other more exotic scenarios. A complete review is beyond the scope of this work and refer the reader to a more comprehensive study on scalar extensions~\cite{Ivanov:2017dad} (and references therein). 

Scalar extensions are proposed in order to solve some of the outstanding issues of the SM although some of them are mainly, or at least also, interesting exercises in model building. The next section (section~\ref{DM}) will be devoted to DM models so here we just state that adding scalar dark matter to the SM is a simple exercise in model building - one needs at least a gauge singlet and a symmetry that will hide the dark from the visible sector.
The most common used is a $\mathbb{Z}_2$ symmetry, attributing to the visible and to dark sector opposite parities. A portal coupling in the Lagrangian~\cite{Patt:2006fw} links the two sectors. 
Another problem that the SM cannot handle is to provide a strong first-order phase transition needed for electroweak baryogenesis as will be discussed in section~\ref{sec:cosmo}. The energy barrier that gives rise to the first-order phase transition can be increased by the addition of new scalar fields. Finally another issue to be solved is the addition of more sources of CP-violation to generate the baryon asymmetry of the universe. Once these issues are taken care of, it is now possible to find models that solve them simultaneously, e.g. with a combination of doublets or doublets and singlets if we consider that only the scalar sector is extended.
The minimal model in terms of number of fields is a 2HDM plus a singlet but several types of 3HDM have also been studied in the literature (see ~\cite{Ivanov:2017dad} for a review on the models).  

Model building can also serve as an exercise to find new possible signatures at the LHC and to stretch our understanding of the world. In the first case we give the example of the well-known Georgi-Machacek model proposed in~\cite{Georgi:1985nv, Chanowitz:1985ug}. The model has two triplets and the usual Higgs doublet. The model features e.g. doubly charged Higgs bosons and tree-level vertices of the type $W^\pm Z H^\mp$, where $H^\mp$ is a singly charged Higgs boson. Finally, the neutral Higgs couplings to the gauge bosons can be larger than the SM-like one.  
A second case is the example of a 3-Higgs doublet models which is CP -conserving, but the coefficients in its scalar potential cannot be made simultaneously real by any basis change. When the Lagrangian in written in terms of physical fields, the extra neutral Higgs bosons have a peculiar property: they are neither CP -even nor CP -odd but, in a well-defined mathematical sense, are “half-odd”~\cite{Ivanov:2015mwl}. These are now known as CP4 models. 

}

\section{Dark matter models}
\label{DM}

Since the rotation of galaxies shows a discrepancy relative to the one predicted with only the visible matter~\cite{Zwicky1933, Rubin:1970zza, Rubin1978} and the term dark matter appeared, physicists have tried to understand both its origin and nature. And although a body of evidence from different sources has been accumulating over the years, it is not clear if DM is a particle, a gravitational effect that could be explained with the modification of Newton's law or some other option that was so far missed by the physics community.

However, once one assumes that DM comes in the form of a particle, at least one new field has to be added to the SM. The only SM particle that could play the role of DM is the neutrino which however seems to be too hot, meaning relativistic, to be in agreement with the structure formation in the universe.  For example, hot DM would not lead to the currently observed large scale structure of the universe
(for a recent review on the problem of neutrinos as DM see~\cite{Buettner:2022xmo}). 

There are no constraints on the DM spin and almost no constraints on its mass. Hence, from the point of view of an extended scalar sector, the simplest way to complete the SM with a DM candidate is to add a real singlet scalar field to the SM field content~\cite{Silveira:1985rk, Burgess:2000yq, Barger:2007im, Guo:2010hq}  via a portal coupling in the Lagrangian, as first proposed in~\cite{Patt:2006fw}. Such a term in the potential has the form $\kappa_{HS} H^\dagger H S^2$, where $\kappa_{HS}$ is the dimensionless coupling that gauges the strength between the dark and the visible sector, $H$ is the Higgs doublet and $S$ is the DM singlet. DM is stabilized by using a symmetry under which the visible and the dark sector have opposite parities. This way the two sectors only communicate via the portal couplings if the DM particle transform trivially under the SM symmetries. The most commonly used symmetry is $\mathbb{Z}_2$ but many others are possible. If for instance the new field from the dark sector has a non-zero isospin, it will also couple to gauge bosons.
The visible sector can be just the SM or some extension thereof to include for instance more sources of CP-violation. The dark sector can include any number of DM candidates, and combinations of singlets, doublets, triplets and more exotic representations have been discussed in the literature.  

There are several mechanisms that are able to reproduce the DM relic density observed today. We focus on the so-called Weakly Interacting Massive Particles (WIMPs) region, with masses in the GeV to TeV ranges. This is the mass region that fits more naturally the idea of extended scalars but more importantly these are the mass ranges that can be probed at the LHC and future colliders. Note that particles with masses below about 1 GeV are effectively massless at the LHC. One of the most popular of the possible mechanisms is freeze-out~\cite{Zeldovich:1965gev, Bertone:2004pz} in which case DM starts in thermal equilibrium with the thermal bath until the annihilation of DM into SM particles stops. 
Another possibility is freeze-in~\cite{Hall:2009bx} in which DM particles are produced via either decay or annihilation of other particles from the visible sector. In this scenario the portal coupling has to be very weak except is some particular cases, and the DM candidates are 
known as  Feebly Interacting Massive Particles (FIMPs). There are a number of proposals that either combine them or add modified versions of the two mechanisms above~\cite{DEramo:2010keq, Hochberg:2014dra, Kuflik:2015isi, DAgnolo:2015ujb, Pappadopulo:2016pkp, Garny:2017rxs, DAgnolo:2020mpt, Smirnov:2020zwf, Fitzpatrick:2020vba, Kramer:2020sbb,
Hryczuk:2021qtz, Bringmann:2021tjr}.

Once a model is proposed, it has to be in agreement with the available experimental data. Let us discuss the most relevant ones on constraining the dark sector, that is, the DM mass and the portal(s) coupling(s). There was a time in the history of the universe where DM relic density reached a value close to the one we measure today  - this is known as DM relic density $\Omega_{\text{DM}}  h^2$ and was measured by the Planck satellite~\cite{Planck:2018vyg} under the assumption of the validity of the Lambda-CDM cosmological model \cite{Peebles:1984ge,Carroll:2000fy,Peebles:2002gy}. All models are severely constrained by this very precise measurement of $\Omega_{\text{DM}}^{\text{obs}}h^2=0.120 \pm 0.001$ from \textsc{Planck}, that fixes a tiny region of allowed points in the portal couplings - DM mass plane. 
Direct detection (DD) bounds are the ones that come from the actual search for events of DM colliding with some atom which in the most recent and precise experiments is liquid Xenon. If no DM is found, they provide constraints in the portal coupling, DM mass plane in the WIMP mass region. The plots presented in this work include bounds from the following DD experiments: XENON1T~\cite{XENON:2018voc}, DarkSide-50~\cite{DarkSide:2018bpj}, PICO-60~\cite{PICO:2019vsc}, CRESST-III~\cite{CRESST:2019jnq}, PandaX-4T~\cite{PandaX-4T:2021bab} and LUX-ZEPLIN (LZ)~\cite{LZ:2022lsv}. Indirect detection of DM also places constraints on the model, although a number of uncertainties make them less reliable than the ones from DD. Also, in the WIMP mass region, they are usually less relevant than the DD ones. 

All bounds are obtained with \texttt{micrOMEGAs 6.0}~\cite{Alguero:2023zol}, except for indirect detection where we have also used  MADHAT~\cite{Boddy:2018qur,Boddy:2019kuw}, with data from dwarf galaxies presented in~\cite{Boddy:2019qak}. Finally collider bounds should also be taken into account. The Higgs could decay into DM particles via the portal coupling. The amount of DM with origin in the Higgs is part of what is called the Higgs invisible width measured by both ATLAS \cite{ATLAS:2023tkt} and CMS \cite{CMS:2025jwz}. These bounds only apply to DM with a mass less than half of that of the Higgs boson, such that an on-shell decay is kinematically allowed. DM can also be produced at colliders alongside a SM particle such as a Higgs or a gauge boson. These are called mono-X events and could become relevant at the high-luminosity stage of the LHC. 

\begin{figure}[h]
    \centering
    \includegraphics[height=0.45\textwidth]{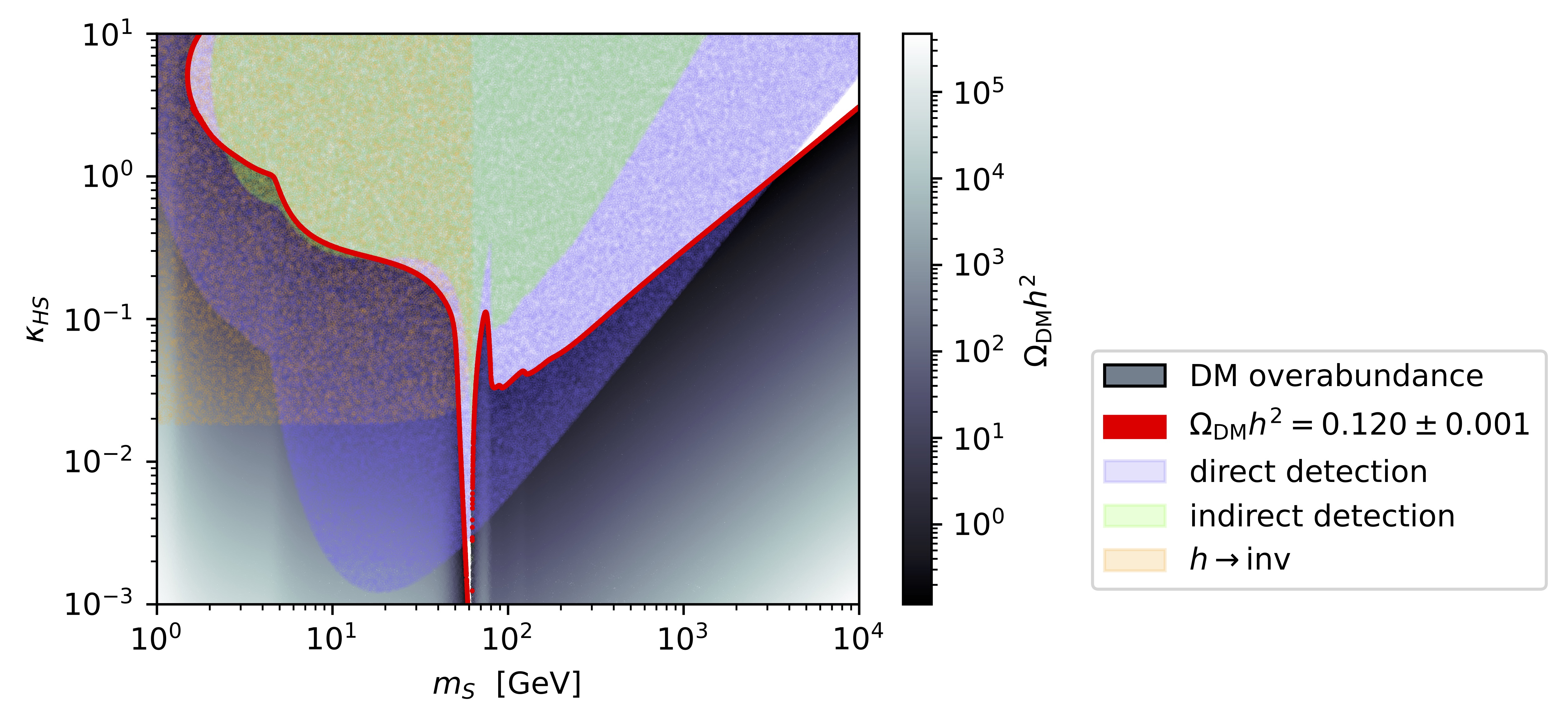}
    \caption{Experimental constraints for the real singlet extension of the SM obtained using \texttt{micrOMEGAs 6.0} for the freeze-out mechanism. The grey, purple, green and orange regions are, respectively, excluded by DM relic (over)density, direct detection, indirect detection and BR($h$$\rightarrow$inv); while the red region corresponds to the observed DM relic density. Figure taken from~\cite{Goncalves:2025snm}.}
    \label{fig:DM1}
\end{figure}

In figure~\ref{fig:DM1} we present the portal coupling $\kappa_{HS}$ as a function of the DM mass when all constraints previously described together with the theoretical constraints that make the model viable and perturbative are considered. This model has only one portal coupling that connects the visible and dark sector and only one DM candidate. The plot was obtained using \texttt{micrOMEGAs 6.0} for a freeze-out DM candidate. 
There are two allowed regions. One where the DM mass is about half the Higgs mass, usually referred to as the resonant region and one starting at a mass of about $3500$ GeV. This value is in the intersection between the line that shows the correct DM relic density curve and the DD bound shown in purple.
The reasons for these two regions are as follows. The freeze-out relic density is proportional to $m_s^2/\kappa_{HS}^2$ while the remaining bounds require small couplings, that can be larger if the masses grow. The exception is the resonant regions where the cross sections for DM depletion are enhanced by the resonance, and therefore allow for smaller couplings.

We have just analyzed the simplest possible scenario of a scalar DM particle. Let us now discuss what is gained with a more complex version of the dark sector. Once you add an extra singlet to the model there are a number of possibilities with very different outcomes. Here we will just briefly discuss three scenarios~\cite{Capucha:2024oaa, Goncalves:2025snm, Capucha:2025iml}. Let us start by noting that when we have more than one DM, the fraction of DM of a single candidate is defined as $\Omega_{\text{DM}}^{S_i}/\Omega_{\text{DM}}^{\text{obs}}$. This is a quantity that can vary between 0 and 1 for each of the $S_i$ candidates and is particularly relevant for direct and indirect detection experiments.  \\

\begin{itemize}
\item \textbf{One more singlet with an independent $\mathbb{Z}_2$ symmetry with freeze-out only} - In this scenario we end up with two DM candidates each one stabilised by its own symmetry. A new region of allowed parameter space opens up because now the DM fraction
is split between the two candidates. The main point here is that the direct detection cross section has to be multiplied by the corresponding DM fraction. If this fraction is very small, the DD bounds will be weaker and the model will be less constrained. Adding more singlets will not open up new parameter space but will allow for a large allowed mass region. 
\\
\item \textbf{One more singlet with just one $\mathbb{Z}_2$ symmetry and freeze-out} - In this scenario we end up with with only one DM candidate. However the allowed region for the DM mass will open up starting from half the Higgs mass, the resonant region, and with no upper bound. The reason for this opening is simple. The model has two portal couplings and one coupling that connects the two fields from the dark sector. In order to have the correct relic density one needs large portal couplings. When there is only one portal coupling this is in conflict with direct detection. However, since in this scenario we have three couplings it is possible to have a small coupling for direct detection and another large coupling to comply with the measured relic density. Thus, the tension between DD and relic density is alleviated.    
\\
\item \textbf{One more singlet with an independent $\mathbb{Z}_2$ symmetry, with freze-out and freeze-in} - In this scenario we end up with two DM candidates, one that freezes-out and one that freezes-in. This is an example of what can be considered a simple way to solve the problem of not having the correct relic density in an extension of the SM. In fact, if a model with a freeze-out DM candidate, has an under abundant relic density, it is always possible to add more DM fields with independent symmetries which freeze-in, to complete the relic density measured experimentally.\\[0.2cm]

\end{itemize}

There are some interesting proposals in the literature that alleviate the tension between direct detection, which needs a small portal coupling, and the measured relic density that needs it to be large. We saw that in the resonant region portal couplings can be small and still be in agreement with the measured relic density due to the enhanced DM annihilation cross section. There are however models, that due to a leading-order cancellation of the direct detection cross section still allow for large portal couplings. These are the pseudo Nambu-Goldstone bosons DM first introduced in~\cite{Gross:2017dan}. This a scenario for which a one-loop calculation is needed for a proper comparison with the experimental results~\cite{Azevedo:2018exj}.

As expected, as more fields are added to the Higgs potential, the models become less and less constrained by experiment. One should note however, that as soon as fields other than singlets are added, for instance $SU(2)$ Isopsin doublets, the dark sector will couple to gauge bosons. Since particles from the dark sector will now couple with gauge couplings, rather than portal couplings with origin in the potential, the interactions between the visible and the dark sector are usually more constrained. The simplest example of such a model is the Inert Doublet Model (IDM), first discussed in \cite{Deshpande:1977rw,Barbieri:2006dq,Cao:2007rm}. This model contains a second doublet that is odd under an additional $\mathbb{Z}_2$ symmetry that remains exact. Therefore, all novel scalars $H,
\,A,\,H^\pm$ can serve as a DM candidate. Although this model was originally proposed because it could render exact relic density, more recently a combination of theoretical and experimental constraints have limited the allowed parameter space in such a way that only a few viable regions remain. See e.g. \cite{Kalinowski:2020rmb} for a recent discussion of the available phase space as well as possible cross sections at current and future collider facilities. As the dark scalars in this scenario only couple to electroweak gauge bosons, typical collider signatures are single or multiple electroweak gauge bosons and missing transverse energy. Corresponding phenomenological studies are abundant, but so far there is no direct LHC search for this model. For a recent study at Higgs factories see e.g. \cite{Bal:2025nbu}.

Finally, we want to briefly mention the 2HDM+a, a 2HDM with an additional pseudoscalar that serves as a portal to the dark sector in the gauge eigenstate. This model has been widely adopted by the LHC experiments as a prime model to investigate DM related signatures, where an overview can be found in \cite{LHCDarkMatterWorkingGroup:2018ufk}. It contains a large number of free parameters, therefore generic searches can only be presented in certain benchmark scenarios. Recent studies by the ATLAS collaboration have e.g. been presented in \cite{ATL-PHYS-PUB-2024-010}.
For a recent review on dark matter see~\cite{Cirelli:2024ssz}.

{
\subsection{CP-violation in a dark sector}

Extensions of the SM are usually also proposed to provide new sources of CP-violation. In the minimal set with one extra doublet and one extra real singlet a model can be build with both CP-violation and a DM candidate. A very interesting scenario with this same set-up is to have secluded dark matter, that is, a dark sector with CP-violation~\cite{Cordero-Cid:2016krd, Azevedo:2018fmj}.

The model in~\cite{Azevedo:2018fmj} has a discrete $\mathbb{Z}_2$ symmetry of the form
\begin{equation}
\Phi_1 \,\rightarrow \,\Phi_1\;\;\;,\;\;\;
\Phi_2 \,\rightarrow \,-\Phi_2\;\;\;,\;\;\;
\Phi_S \,\rightarrow \, -\Phi_S\,.
\label{eq:z2z2}    
\end{equation}
and the most general scalar potential invariant under $SU(2)\times U(1)$ and the above symmetries is given by
\begin{eqnarray}
V &=& m_{11}^2 |\Phi_1|^2 \,+\, m_{22}^2 |\Phi_2|^2 \,+\, \frac{1}{2} m^2_S \Phi_S^2
\, +\, \left(A \Phi_1^\dagger\Phi_2 \Phi_S \,+\,h.c.\right)
\nonumber \\ & &
\,+\, \frac{1}{2} \lambda_1 |\Phi_1|^4
\,+\, \frac{1}{2} \lambda_2 |\Phi_2|^4
\,+\, \lambda_3 |\Phi_1|^2 |\Phi_2|^2
\,+\, \lambda_4 |\Phi_1^\dagger\Phi_2|^2\,+\,
\frac{1}{2} \lambda_5 \left[\left( \Phi_1^\dagger\Phi_2 \right)^2 + h.c. \right] \nonumber \\ & &
\,+\,\frac{1}{4} \lambda_6 \Phi_S^4 \,+\,
\frac{1}{2}\lambda_7 |\Phi_1|^2 \Phi_S^2 \,+\,
 \frac{1}{2} \lambda_8 |\Phi_2|^2 \Phi_S^2 \,,
\label{eq:pot}
\end{eqnarray}
where, with the exception of $A$, all parameters in the
potential are real. After rotating to the mass eigenstates we end up
with the 125 GeV Higgs and three DM particles we call $h_1$, $h_2$ and $h_3$. The DM candidate is the lightest of the three.

As for the Yukawa sector, we consider all fermion fields are
even under this symmetry. Therefore, only  $\Phi_1$ couples to fermions, and the Yukawa Lagrangian is 
\begin{equation}
-{\cal L}_Y\,=\, \lambda_t \bar{Q}_L \tilde{\Phi}_1 t_R\,+\,\lambda_b \bar{Q}_L \Phi_1 b_R
\,+\,\lambda_\tau \bar{L}_L \Phi_1 \tau_R\,+\,\dots
\label{eq:yuk}
\end{equation}
where we have only written the terms corresponding to the third generation of fermions, with the
Yukawa terms for the remaining generations taking an analogous form. The
left-handed doublets for quarks and leptons are denoted by $Q_L$ and
$L_L$, respectively; $t_R$, $b_R$ and $\tau_R$ are the
right-handed top, bottom and $\tau$ fields; and $\tilde{\Phi}_1$ is the
charge conjugate of the doublet $\Phi_1$. 

The fact that CP-violation is located in the dark scalar sector poses automatically the question whether this effect could be observed at colliders.  A Lorentz structure analysis
of the $ZZZ$ vertex~\cite{Hagiwara:1986vm,Gounaris:1999kf, Baur:2000ae, Grzadkowski:2016lpv}, shows that it can be reduced to two form factors if one assumes two on-shell $Z$ bosons. With these assumptions, the $ZZZ$ vertex function can be written as
\begin{equation}
e\Gamma^{\alpha\beta\mu}_{ZZZ} \,=\,i\,e\,\frac{p_1^2 - m^2_Z}{m^2_Z} \left[f^Z_4 \left(p_1^\alpha g^{\mu\beta} +
p_1^\beta g^{\mu\alpha}\right)\,+\,f^Z_5 \epsilon^{\mu\alpha\beta\rho}
\left(p_2-p_3\right)_\rho\right] \;,
\label{eq:vert}
\end{equation}
where $e$ is the electric charge, $p_1$ is the 4-momentum of the off-shell $Z$ boson, $p_2$ and $p_3$ those of the remaining (on-shell) $Z$ bosons.

\begin{figure}[h!]
\begin{center}
\includegraphics[height=3cm,angle=0]{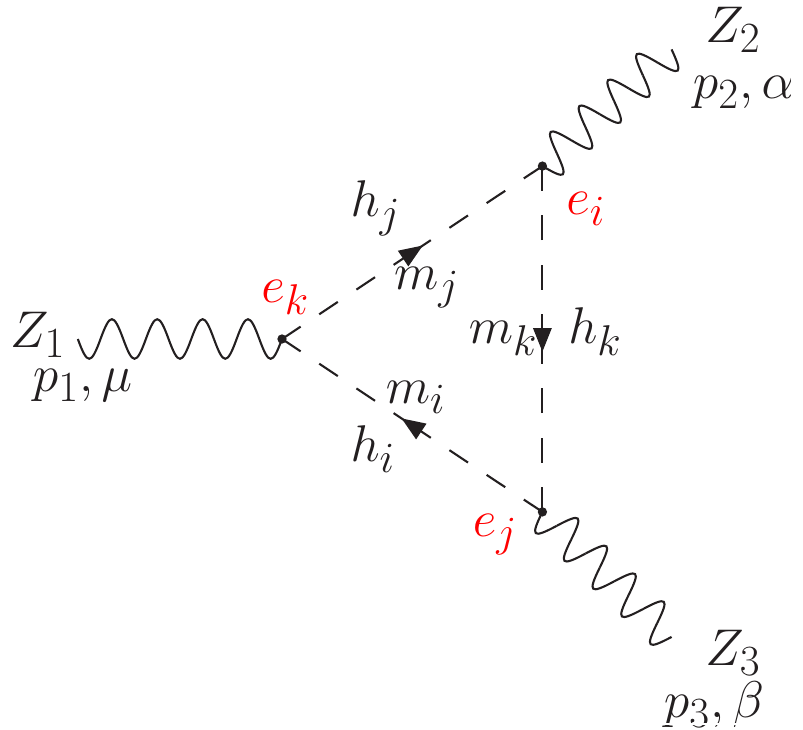}
\caption{\label{fig:diagram} Feynman diagram contributing to the CP violating form factor $f^Z_4$. Note that $i \neq j \neq k$ and $e_n$, $n=i,j,k$ are the couplings.
}
\end{center}
\end{figure}
The form factor $f_4^Z$ is CP-violating while $f_5^Z$ is CP-conserving. CP-violation can thus be probed by the process shown in diagram in figure~\ref{fig:diagram}. The presence of a non-zero value of $f_4^Z$ is only guaranteed if there are three different neutral scalars circulating in the loop, as shown in~\cite{Grzadkowski:2016lpv, Belusca-Maito:2017iob} for the C2HDM - the 2-Higgs doublet model with explicit CP-violation. In these models, the maximum values of 
$f^Z_4$ are typically below $10^{-5}$. ATLAS~\cite{ATLAS:2023zrv} and CMS~\cite{CMS:2020gtj} set bounds on $f^Z_4$ using the process $pp \to ZZ$ with bounds of the order of $10^{-3}$. It is therefore possible that these models could be probed in the LHC high luminosity run.

}

\bigskip

\section{Current experimental status}\label{sec:current}
The LHC experiments are pursuing a vast program of experimental searches for new physics scenarios. It is therefore impossible to list all possible searches and bounds in this section. We therefore only give some sample results from current LHC searches and refer the reader to \cite{atlpub, cmspub} 
for more details.

Most of the new physics scenarios contain a large number of additional model parameters. In such scenarios, it is difficult to display the current constraints in 2-dimensional planes. A typical strategy in this case is then to fix nearly all but two free parameters, which leads to relatively simple exclusion contours. However, in such cases care must be taken that the displayed bounds depend on the assumptions and can differ significantly if these are not fulfilled. 

Another way of presenting experimental limits is to show bounds on rates as a function of one or more of the physical parameters of the model. As an example, in figure~\ref{fig:singlets} we show the bounds on resonance-enhanced di-Higgs searches as well as bounds on the decay of a heavy scalar into two lighter ones, $h_3\,\rightarrow\,h_1\,h_2$ production and decay in various channels from the CMS collaboration.

{
\begin{figure}[htb!]
\begin{center}
\includegraphics[width=0.46\textwidth]{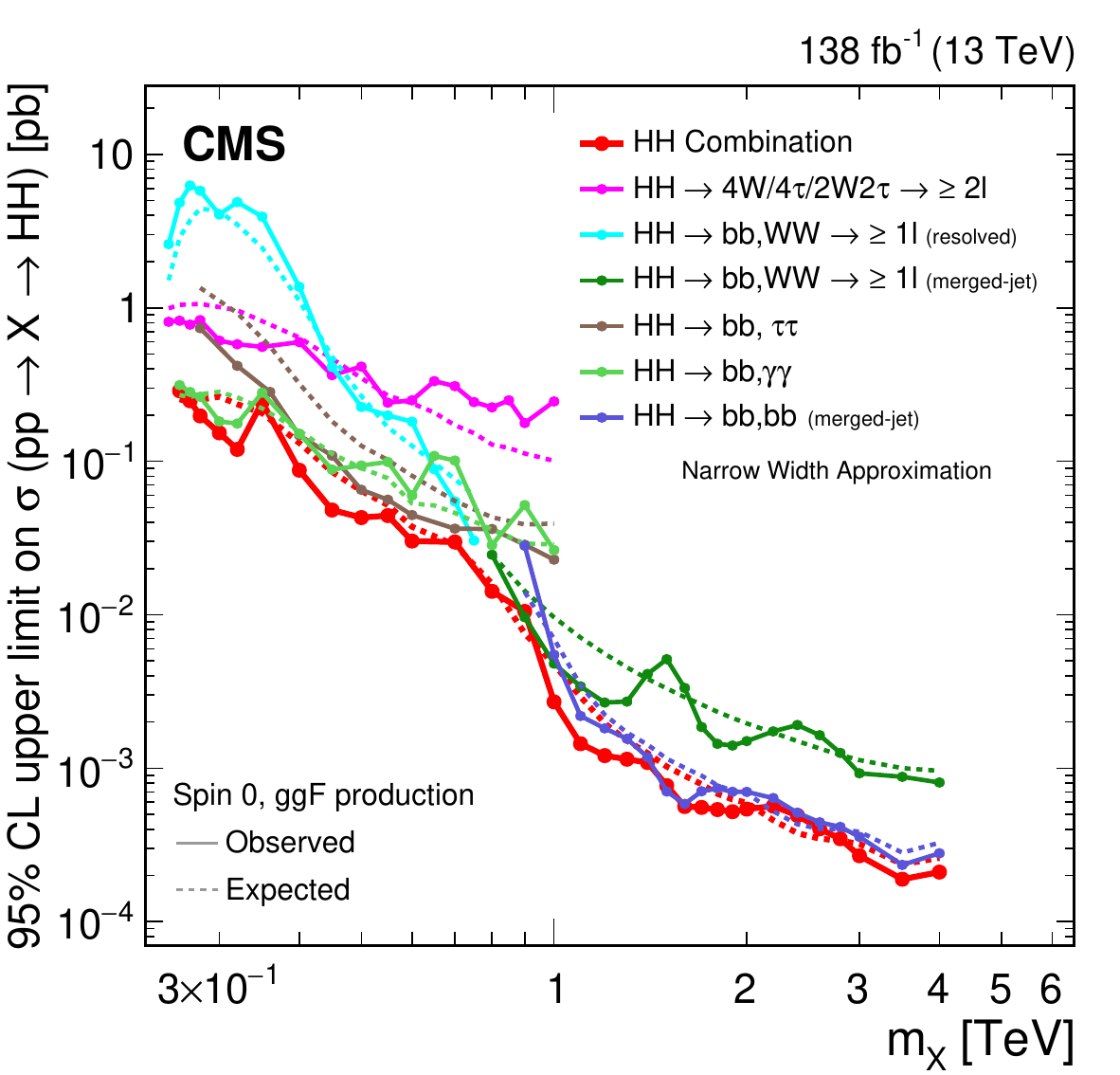}
\includegraphics[width=0.3\textwidth]{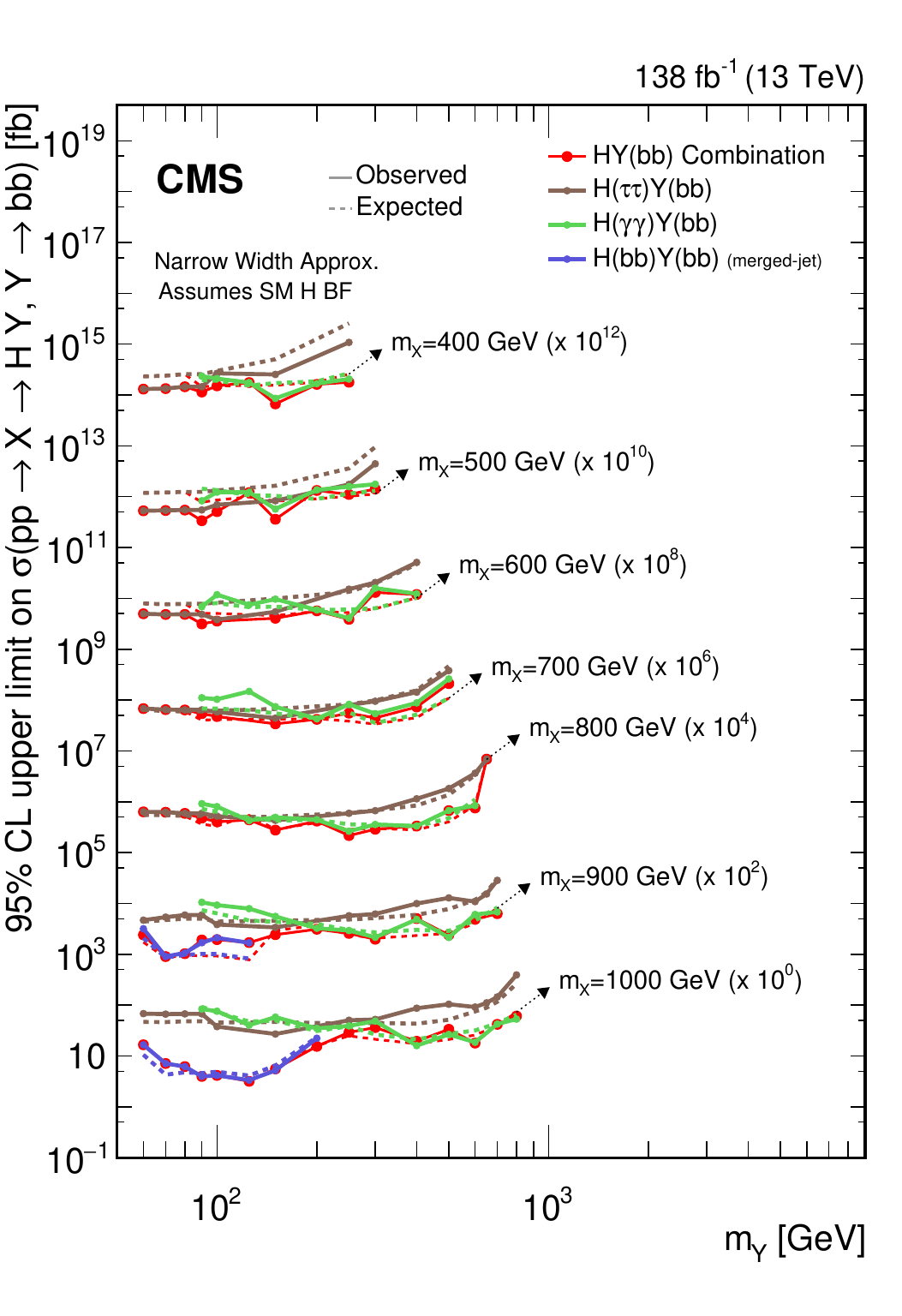}
\caption{\label{fig:singlets} Current experimental bounds on production times decay in the scalar sector, from the CMS Run 2 combination \cite{CMS:2024phk}. {\sl (Left):} from heavy scalar resonances decaying into di-Higgs final states. {\sl (Right):} for processes $X\,\rightarrow\,H\,Y$ ($h_3\,\rightarrow\,h_1\,h_2$ in our notation), for various final states.}
\end{center}
\end{figure}

In figure~\ref{fig:type1atl}, we exemplarily show recent summary plots for type 1 2HDMs as well as the hMSSM, a supersymmetric variant with a type 2 2HDM sector, from the ATLAS collaboration, taken from \cite{ATL-PHYS-PUB-2024-008}. Specifics about the hMSSM setup used here can be found in \cite{Bagnaschi:2039911} (see also \cite{Maiani:2012ij,Djouadi:2013uqa,Djouadi:2015jea} for the original model definition).

\begin{figure}[htb!]
\begin{center}
    \includegraphics[width=0.45\textwidth]{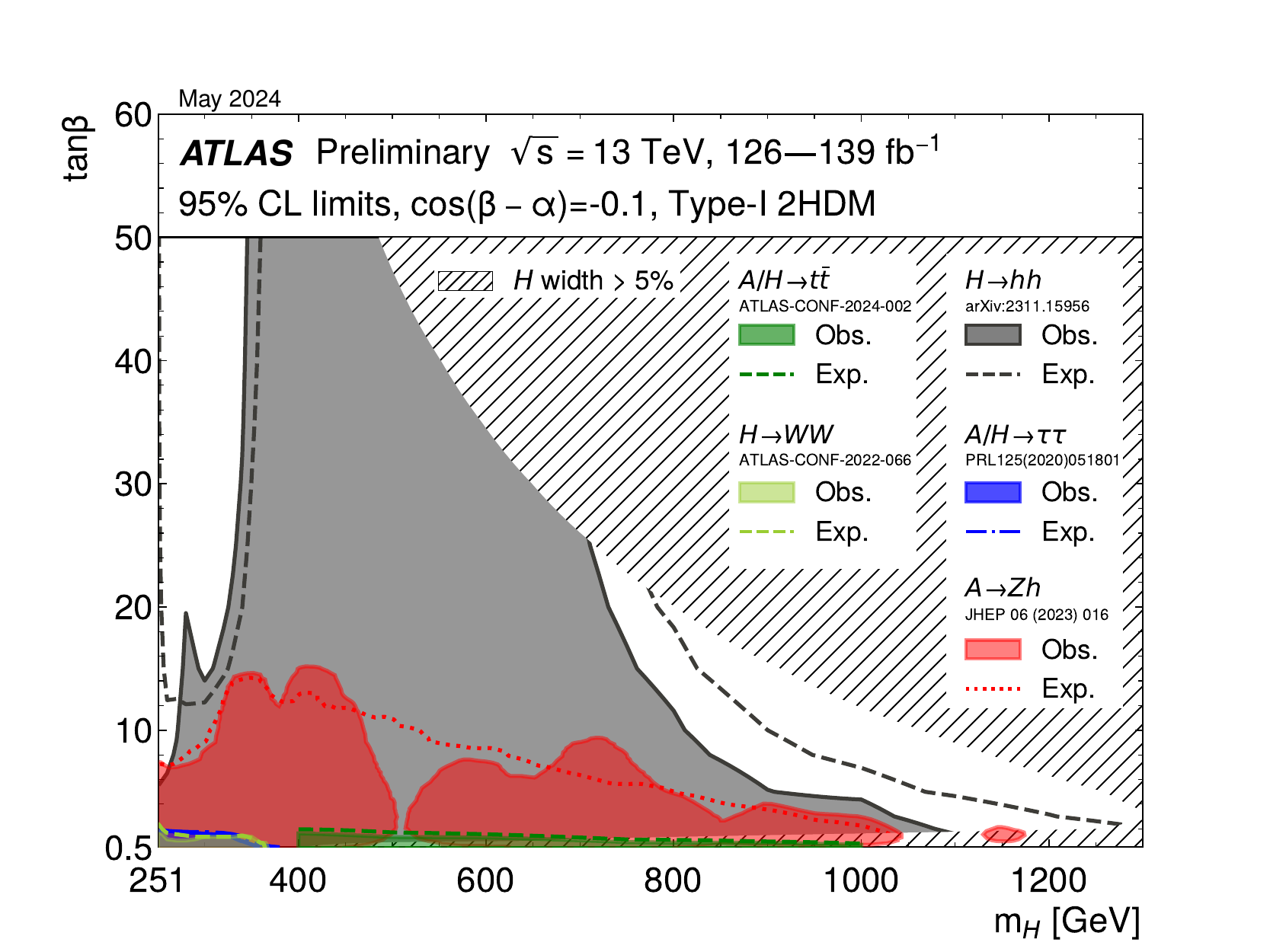}
     \includegraphics[width=0.45\textwidth]{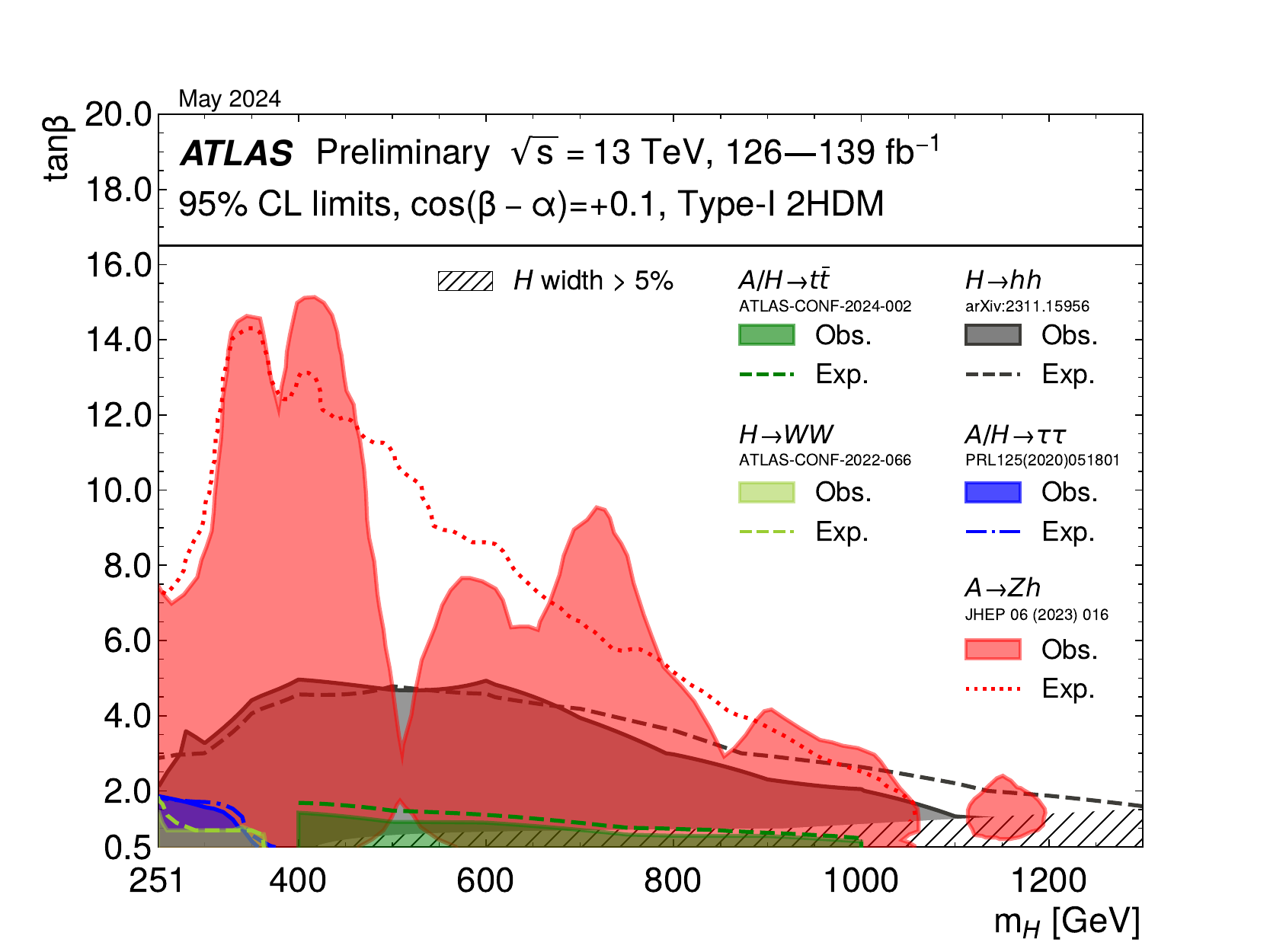}\\
      \includegraphics[width=0.45\textwidth]{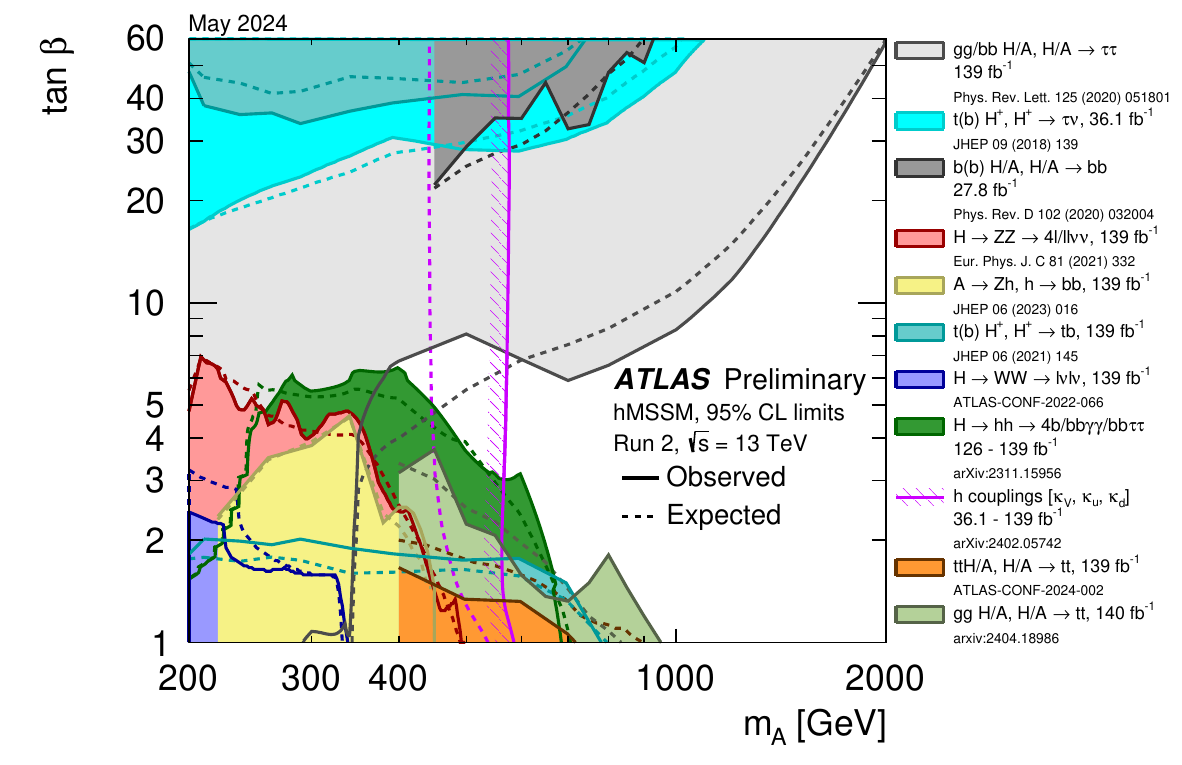}
     \caption{\label{fig:type1atl}  {\sl Top:} Summary plots for type I 2HDM scenarios from the ATLAS collaboration, taken from \cite{ATL-PHYS-PUB-2024-008}, in the $\lb m_H,\,\tan\be\rb$ plane. {\sl (Left)} for $\cos\lb \be-\al\rb\,=\,-0.01$ and {\sl (right)} for $\cos\lb \be-\al\rb=+0.01$. Shown are observed and expected bounds including several full Run 2 searches. A limit on the total width of the new scalars is also included.
     {\sl Bottom:} Bounds in the hMSSM parameter space, taken from the same reference.}
\end{center}
\end{figure}

Obviously, one can also obtain bounds from using some of the publicly available tools discussed in section~\ref{sec:tools}. In figure~\ref{fig:2hdm_htools}, we display bounds on two variants of the 2HDM obtained using thdmtools~\cite{Biekotter:2023eil}/ HiggsTools~\cite{Bahl:2022igd}.  The figure is taken from \cite{Robens:2025tew}. Note that for these plots all additional scalar masses were set equal, while all other parameters were freely floating.

\begin{figure}[htb!]
        \begin{center}
            
            \includegraphics[width=0.45\textwidth]{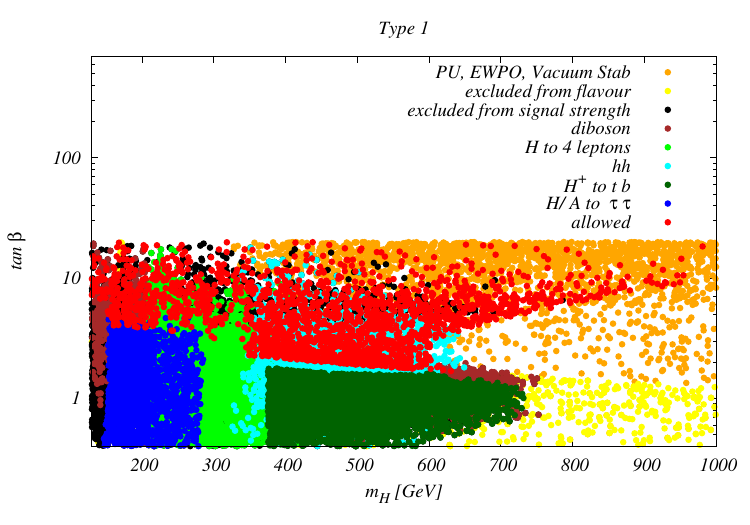}
            \includegraphics[width=0.45\textwidth]{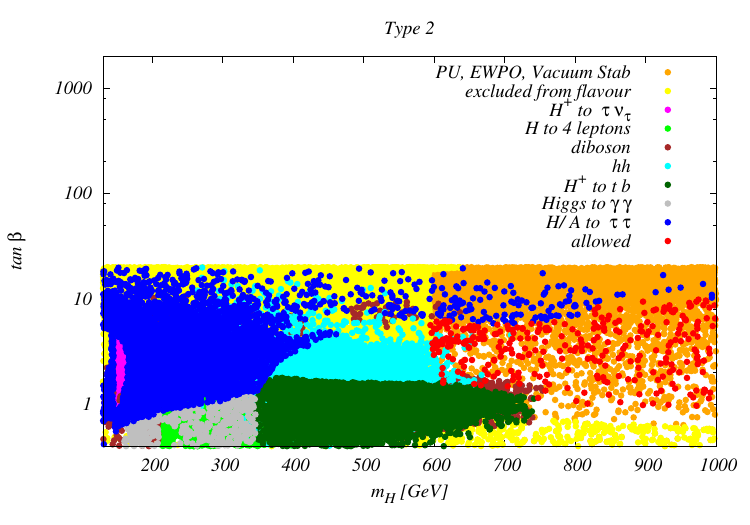}
            \caption{\label{fig:2hdm_htools} Current experimental status of type 1 {\sl (left)} and type 2 {\sl (right)} 2HDMs, obtained through thdmtools. Taken from \cite{Robens:2025tew}. See text for details.}
        \end{center}
\end{figure}

In figure~\ref{fig:2hdma} we display current bounds for the 2HDMa, a 2-Higgs doublet model that contains an additional pseudoscalar field that serves as a portal to the dark sector (see e.g. ~\cite{Ipek:2014gua,No:2015xqa,Goncalves:2016iyg,Bauer:2017ota,Tunney:2017yfp} for original discussions and \cite{LHCDarkMatterWorkingGroup:2018ufk,Robens:2021lov,Argyropoulos:2022ezr,Arcadi:2022lpp,Argyropoulos:2024yxo} for more recent work).  
This model, although containing a fermionic dark matter candidate, is of interest as it currently presents one of the standard scenarios investigated by the LHC experiments where all visible new physics signatures stem from an enhanced scalar sector. It additionally contains 12 novel parameters in the scalar sector, and therefore serves as a prime example of the above discussion that in general bounds that are performed in 2-dimensional planes with all other parameters fixed do not necessarily translate to other regions in parameter space where the remaining parameters have been set to other values. 
The plots in figure~\ref{fig:2hdma} show the current bounds as given in \cite{ATL-PHYS-PUB-2024-010}. We display one example for constraints in the "standard" $\lb m_a,\,m_A \rb$ plane and one in the $\lb m_a,\,m_\chi\rb$ plane, where $\chi$ here denotes the fermionic DM candidate.

\begin{figure}[h!]
        \begin{center}
        \includegraphics[width=0.45\textwidth]{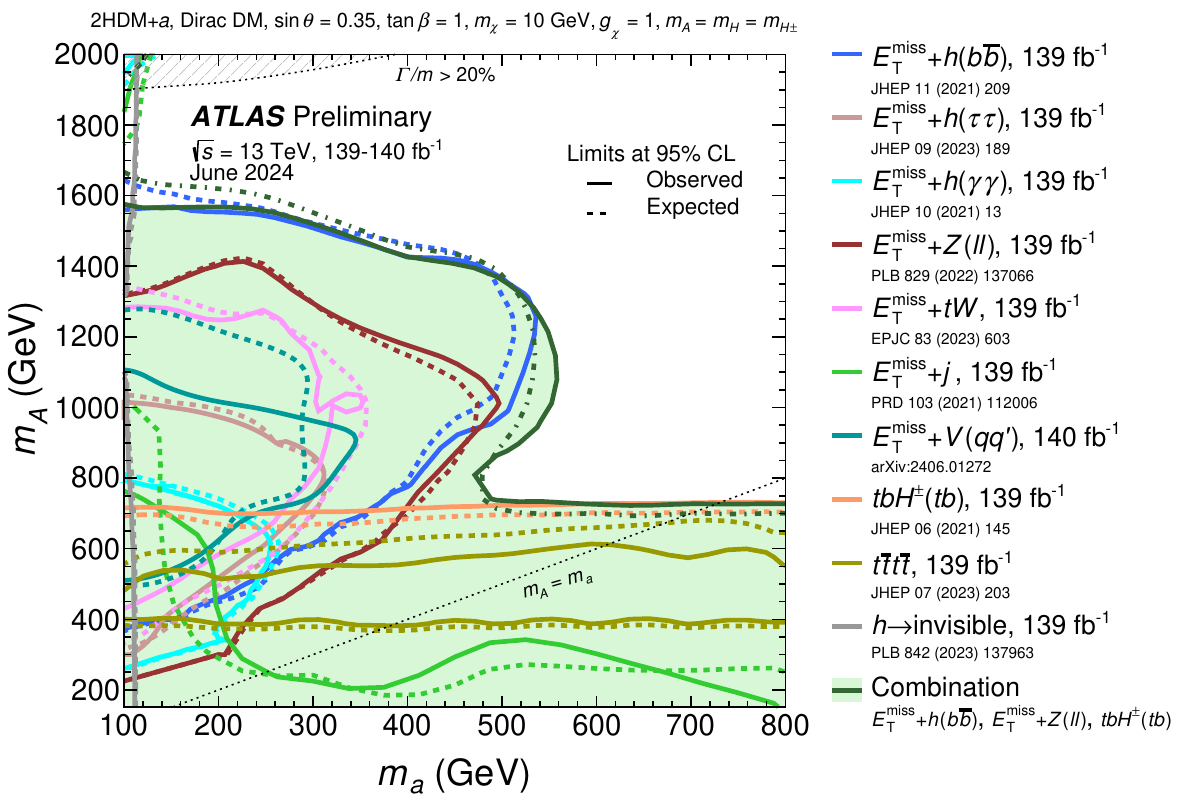}
        \includegraphics[width=0.45\textwidth]{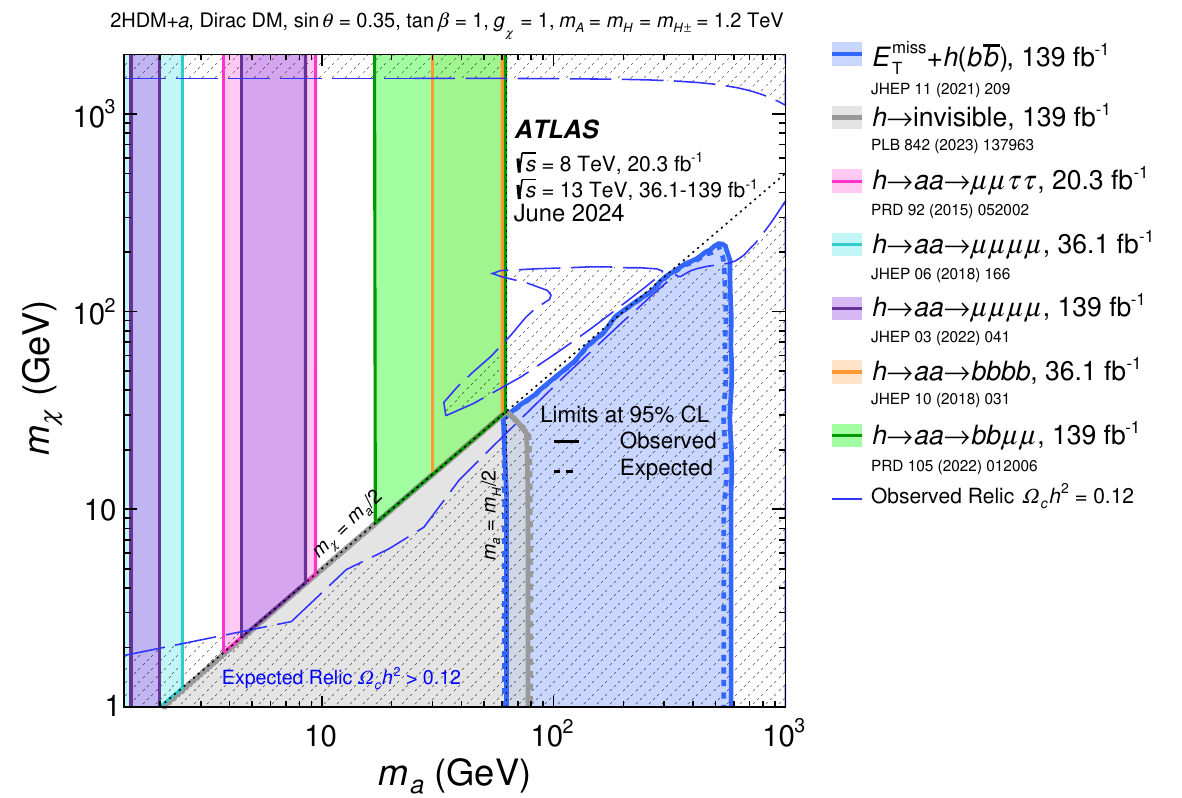}
        \caption{\label{fig:2hdma} Exclusion bounds in the 2HMDa, from combinations as well as various single searches, in the $\lb m_a,\,m_A\rb$ {\sl (left)} and $\lb m_a,\,m_\chi\rb$ {\sl (right)} plane. All other free parameters are fixed here. Taken from \cite{ATL-PHYS-PUB-2024-010}.}
        \end{center}
    \end{figure}

Apart from current LHC searches, there have of course been quite a few direct search results from previous collider experiments.
For example, the LEP experiments have provided a large number of search bounds that are summarized in \cite{OPAL:2002ifx,ALEPH:2006tnd}. In particular for low mass scalars with $m_{h_i}\,\leq\,120\,\GeV$, bounds from LEP still pose the strongest constraints in some regions of parameter space.
Finally, there is a model-independent bound on the mass of charged scalars for models containing a dark matter candidate. For such models, rederived LEP bounds \cite{OPAL:2003nhx,OPAL:2003wxm} render \cite{Pierce:2007ut} $m_{H^\pm}\,\geq\,70-90\,\GeV$. A summary of the physics results obtained at the Tevatron by the CDF and D0 experiments can be found in \cite{Bandurin:2014bhr}.

}


\section{Future collider options}

{
Apart from current experimental data, it is of course instructive to consider the reach of possible future collider experiments for models with extended scalar sectors. Given the length of this article, we can again only give a short overview here and refer the reader to the literature for further details. A very detailed discussion can e.g. be found in the Snowmass Energy Frontier report \cite{Narain:2022qud}.

Both for the Snowmass process in the US as well as the currently ongoing European strategy for particle physics (see e.g. \cite{snowmass, eustrat}), one important topic is to decide about the financing and building of possible future accelerators as well as a corresponding priority list if needed. At the moment, the following options are under discussion with different development stages:
\begin{itemize} 
\item{}High-luminosity LHC (HL-LHC): this is a proton-proton machine that corresponds on an upgrade of the current LHC. It is approved and will start at CERN in 2030 (see e.g. \cite{highlumiwww}), with a center-of-mass energy of 13.6-14 \TeV. The main goal of the HL-LHC is to provide access to some new physics scenarios that currently suffer from low statistics. A detailed discussion can be found in \cite{Cepeda:2019klc},
and some projections are available from the LHC experiments considering reach of novel scalar searches \cite{CMS:2025hfp}.
\item{}$e^+e^-$ machines with relatively low center of mass energies $\leq\,1\TeV$: %
such colliders, also labeled "Higgs factories" 
have received high-priority in the last community efforts both in Europe and the US. A detailed discussion of the physics prospects can be found in \cite{deBlas:2024bmz}. These machines can be designed as either circular or linear accelerators. 
Although the major focus is currently on center-of-mass energies of about 240-250 \GeV, most of them are also designed to run at 90 \GeV (Z-pole), 340-360 \GeV ($t\bar{t}$ threshold), and 160 \GeV ($WW$ threshold) center-of-mass energies 
(see e.g. \cite{elbatalk}). 
The linear colliders additionally allow upgrades to higher center-of-mass energies, e.g. to 500 \GeV or 1 \TeV. Currently under discussion are the FCC-ee at CERN \cite{FCC:2025lpp,FCC:2025uan,FCC:2025jtd}, a linear collider option in Asia or Europe (ILC) \cite{LinearCollider:2025lya,LinearColliderVision:2025hlt,ilcwww}, as well as a circular collider option in China (CEPC) \cite{CEPCStudyGroup:2023quu}. 

\item{}$e^+e^-$ at higher energies: there is also some discussion on $e^+e^-$ machines at higher center-of-mass energies, with a prominent example being CLIC or any other linear machine going up to (multi-) TeV range (see e.g. discussions in \cite{Adli:2025swq,LinearColliderVision:2025hlt}). Although these are currently not counted as the primary target, these collider options can be of interest as several new physics scenarios now already rule out low-mass novel particles and therefore more energy is required. As an example, we have type 2 2HDMs where a combination of flavour and other constraints require mass scales $\geq\,500-600\,\GeV$ for the additional scalars (see e.g. discussion in \cite{Robens:2021lov,Robens:2022jrs}).
\item{}Muon colliders: there has recently been a renewed interest in muon colliders, with a specific focus on \TeV-scale center-of-mass energies (see e.g. \cite{Black:2022cth,Accettura:2023ked,InternationalMuonCollider:2025sys} for recent community efforts). A major advantage of these colliders would be that they reach an energy regime where the electroweak gauge bosons can be considered to be effectively massless, therefore leading to log-enhanced emissions of these bosons in vector-boson-fusion type reactions of the form
\begin{\eqn*}
\mu^+\,\mu^-\,\rightarrow\,\nu_\mu\,\bar{\nu}_\mu\,X
\end{\eqn*}
via a VBF-type type topology\footnote{This is for $W$-boson emission. Emission of $Z$ bosons follows analogously.}. Muon colliders would in principle combine the clean environment of lepton colliders with advantages of going to high-energy regimes. Currently however the realization of such colliders is still a topic of investigation.
\item{}High energy hadron colliders (FCC-hh): finally, very high-energy hadron colliders are also of interest and under discussion. A prominent example for this would be e.g. the FCC-hh, which could be a follow-up project of a new ee collider built around the CERN site \cite{FCC:2025lpp,FCC:2025uan,FCC:2025jtd}. This collider is currently designed to reach center of mass energies of around 100 \TeV. A prime application for this in the scalar sector would be the determination of the triple and quartic scalar couplings, where new physics is taken into account in the form of coupling modifications that can e.g. arise from new physics contributions in higher-order corrections to the Higgs self-couplings, see e.g. discussions in \cite{Abouabid:2024gms} and references therein\footnote{Some recent work on models allowing for large self-coupling enhancements can be found e.g. in \cite{Kanemura:2002vm,Kanemura:2004mg,Braathen:2019pxr,Braathen:2019zoh,Braathen:2020vwo,Bahl:2022jnx,Bahl:2023eau,Falaki:2023tyd,Aiko:2023xui,Braathen:2025qxf}.}.
\end{itemize}

As discussed, for brevity reasons we cannot comment on all possible collider types and new physics options in detail. We here therefore discuss some scenarios exemplarily and refer the reader to the literature (see e.g. presentations in \cite{zimmertalk, bernarditalk}) for further details.

We start with the discussion of Higgs factories, with a center-of-mass energy of around 240-250 \GeV. For such scenarios, within the current European Committee for Future Accelerators (ECFA) effort \cite{ecfa}  a list of prime targets has been identified that should be investigated from the phenomenological viewpoint, see e.g. \cite{deBlas:2024bmz} for a discussion or \cite{Robens:2022zgk,Robens:2025kaa} for (recent) overviews/ updates.
The dominant production mode at $e^+e^-$ colliders of this center of mass energy is the so-called scalar-strahlung process
\begin{\eqn}\label{eq:scalars}
e^+e^-\,\rightarrow\,Z^*\,\rightarrow\,Z\,h_i,
\end{\eqn}
where $h_i$ denotes a scalar of any mass. Searches for this final state have already been performed by the LEP experiments, see e.g. \cite{OPAL:2002ifx}. Typically such searches considered $b\,\bar{b}$ final states for the scalar and leptonic decay modes for the $Z$-boson. Limits can then be obtained explicitly using the scalar decay products or just using the recoil of the $Z$-boson. In \cite{Drechsel:2018mgd}, a sensitivity study was performed using both methods for additional SM-like scalars in a mass range reachable at a 250 \GeV center-of-mass energy ILC. We display the results of this study in figure~\ref{fig:extrasee}.
  \begin{figure}[htb!]
        \begin{center}     
\begin{minipage}{0.32\textwidth}
        \includegraphics[width=\textwidth, angle=-90]{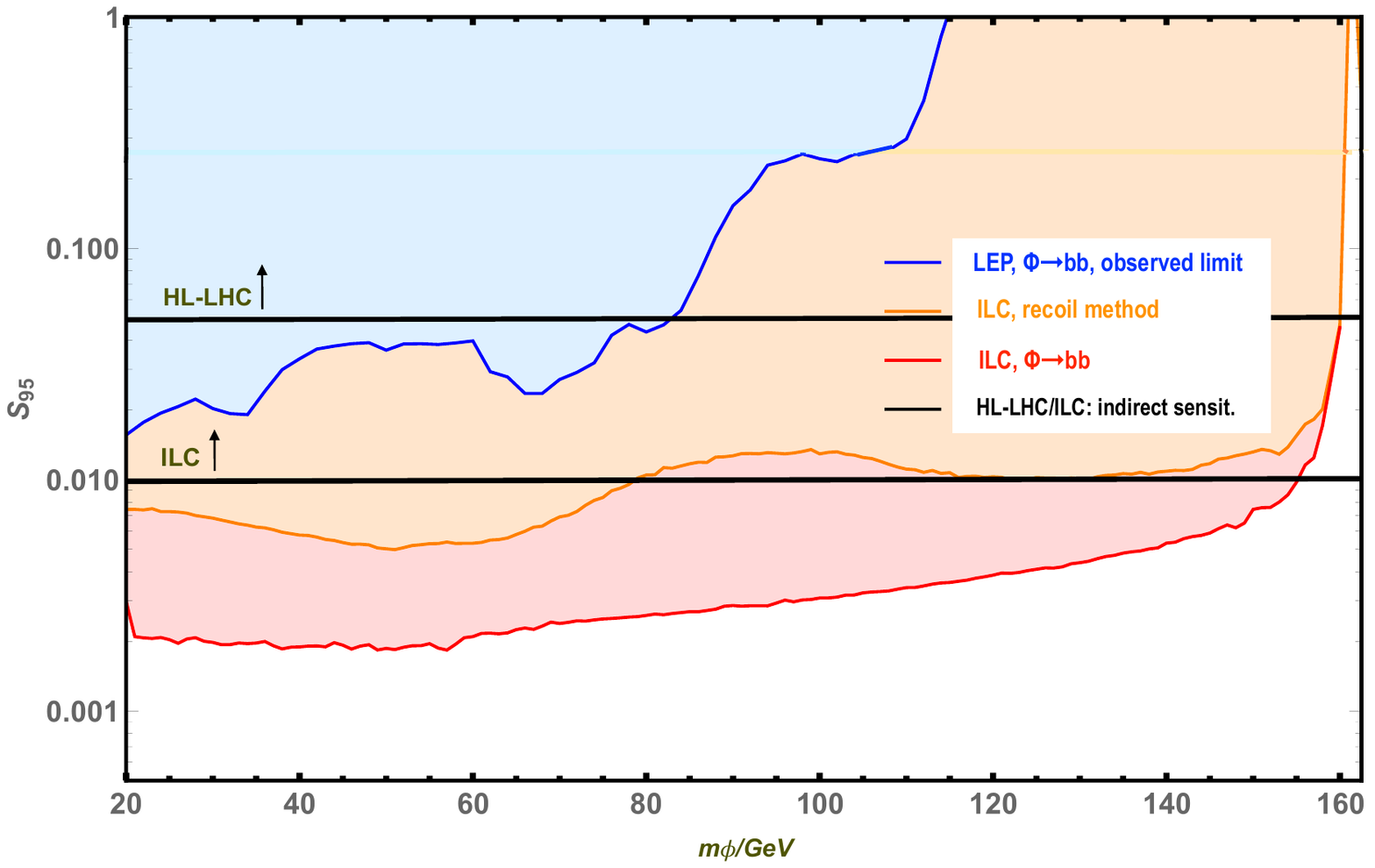}
        \end{minipage}
        \hspace{25mm}
        \begin{minipage}{0.45\textwidth}
        \includegraphics[width=\textwidth]{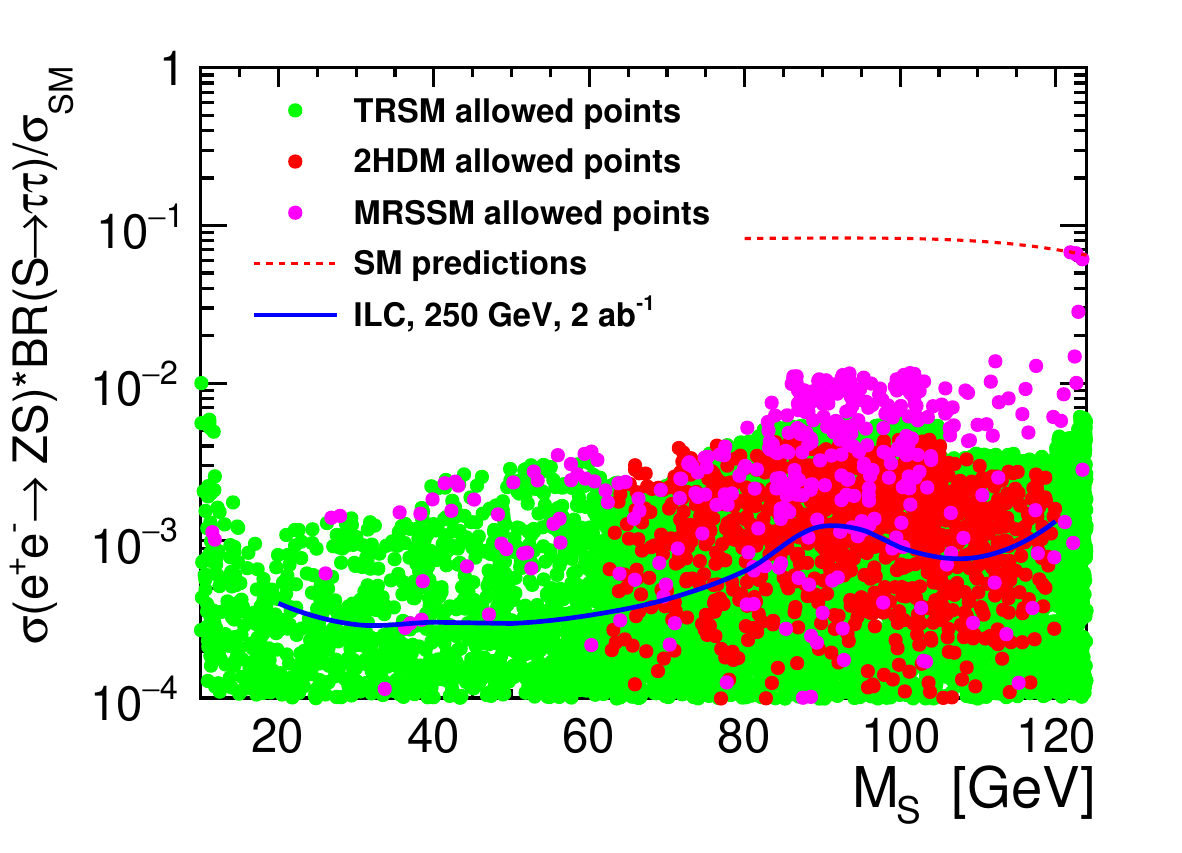}
        \end{minipage}
        \caption{\label{fig:extrasee} {\sl Left:} Projected sensitivity of a 250 \GeV~ LHC for the scalar-strahlung process using both $b\,\bar{b}$ final states as well as the recoil method, taken from \cite{Drechsel:2018mgd}. The quantity $S_{95}$ refers to the cross section that is still compatible with a background only hypothesis at $95\%$ confidence level and can be reinterpreted as a limit on the squared rescaling of the coupling with respect to the SM value at that mass. {\sl Right:} Comparison of the scalar-strahlung with $\tau^+\tau^-$ final states with various new physics scenarios, taken from \cite{Altmann:2025feg}. }
  \end{center}
    \end{figure}

One of the channels that was identified as missing in \cite{deBlas:2024bmz} was the scalar-strahlung process (\ref{eq:scalars}) with a $\tau^+\,\tau^-$ final state for the novel scalar, now available with the study performed in \cite{Brudnowski:2024iiu}. In figure~\ref{fig:extrasee} we present the reach of such a search and compare it to the available parameter space in various new physics scenarios. The figure is taken from~\cite{Altmann:2025feg} and was first presented in~\cite{Robens:2025tew}.
Additional recent discussions of such processes can e.g. be found in~\cite{Berggren:2024aga}.

For the high-luminosity LHC, ATLAS and CMS have recently provided updated projections for various processes in \cite{CMS:2025hfp}\footnote{See also \cite{CMS:2024phk} for projections for specific scenarios from CMS.}. We show the projections for a simple scalar extension, as discussed in section \ref{sec:singlets}, without a $\mathbb{Z}_2$ symmetry such that the number of free parameters now amounts to 5. In figure~\ref{fig:hllhc}, we display the combined projection from ATLAS and CMS in two significant two-dimensional planes. Note that this model also allows for a strong first-order electroweak phase transition (see also section \ref{sec:cosmo}), with the according regions displayed as well. See the original reference for further details.

\begin{figure}[h!]
\begin{center}
\includegraphics[width=0.45\textwidth]{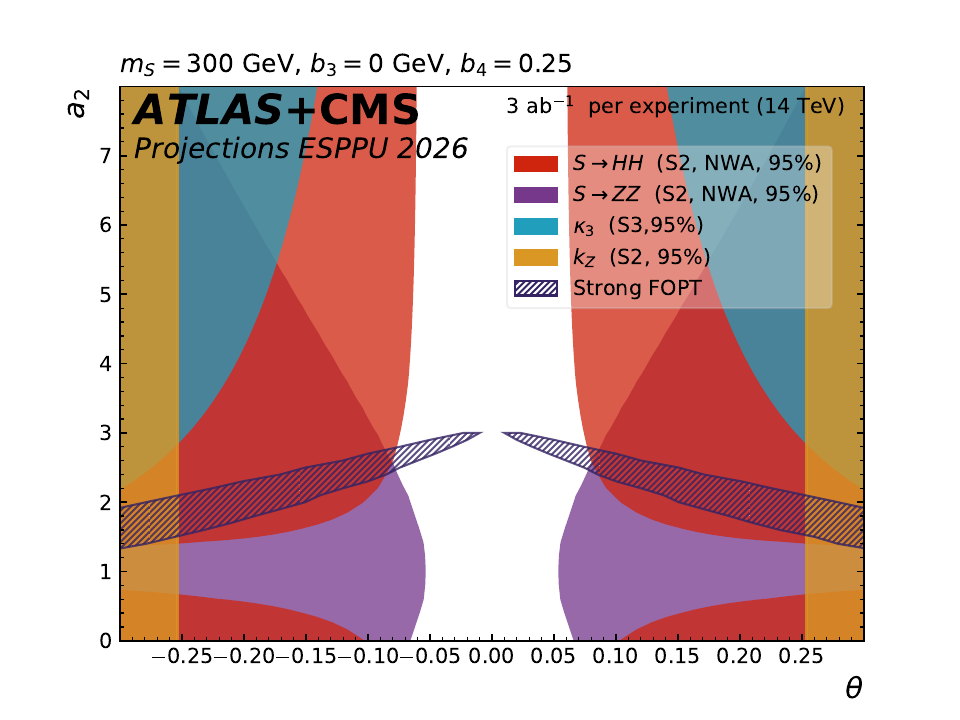}
\includegraphics[width=0.45\textwidth]{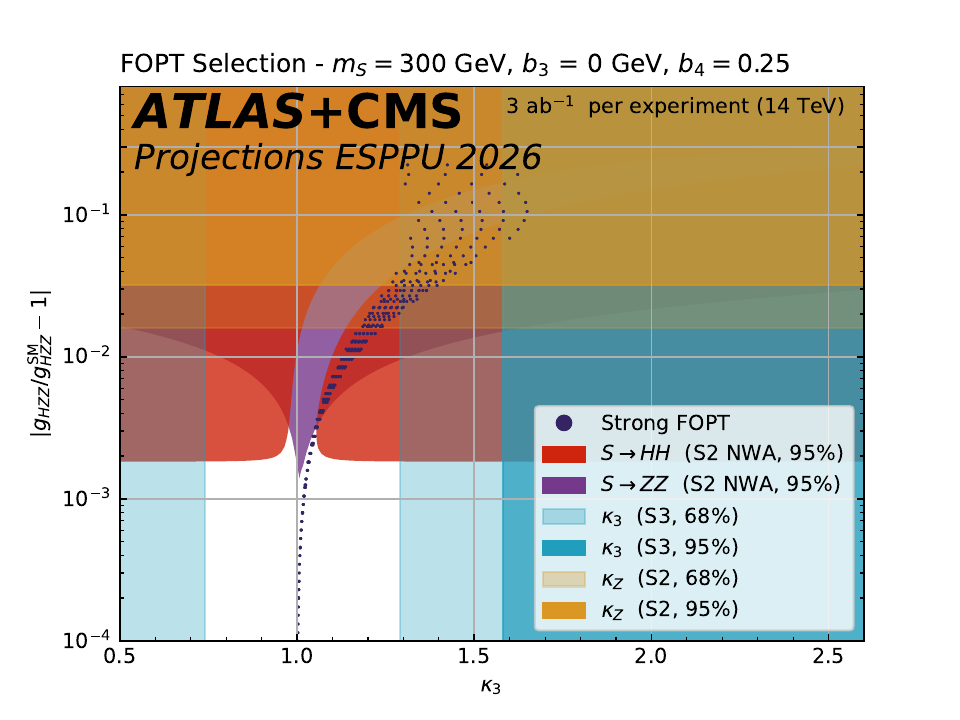}
\end{center}
\caption{\label{fig:hllhc} Combined projection of ATLAS and CMS for a real singlet extension without a $\mathbb{Z}_2$ symmetry, taken from \cite{CMS:2025hfp}. {\sl Left:} in the $\lb \theta,\,a_2 \rb$ plane. {\sl Right:}  $\kappa_3$ on the x-axis, and the y-axis showing the relative deviation of the Higgs boson coupling to $ZZ$ with respect to the same coupling in the SM. Displayed are various search predictions, including predictions on the accuracy of the $\kappa_{3/Z}$ coupling modifiers. $S_{2/3}$ refer to different assumptions for systematic uncertainties (see original reference for details). Shown are also points that allow for a strong first-order electroweak phase transition. The mixing angle between gauge and mass eigenstates is given by $\theta$, and $a_i,\,b_i$ are additional potential parameters.}
\end{figure}

Finally, one can investigate the discovery potential of a possible high-energy muon collider. Such proposals have recently seen a lot of activity. For simplicity, we again concentrate on a model that extends the SM scalar sector by an additional singlet.

In \cite{Accettura:2023ked}, a projection plot was presented for the accuracy that can be reached on the mixing angle of the additional singlet from various possible future collider scenarios, as a function of the additional scalar mass (the work is partially based on \cite{Buttazzo:2018qqp,EuropeanStrategyforParticlePhysicsPreparatoryGroup:2019qin}). We show this plot in figure~\ref{fig:singletmu}. Indirect constraints on the mixing angle here stem from the assumed accuracy on the measurement of the 125 \GeV~ scalar couplings. Direct searches are using the vector-boson-fusion channel for the singlet scenario.

 \begin{figure}[htb!]
        \begin{center}
        \includegraphics[width=0.5\textwidth]{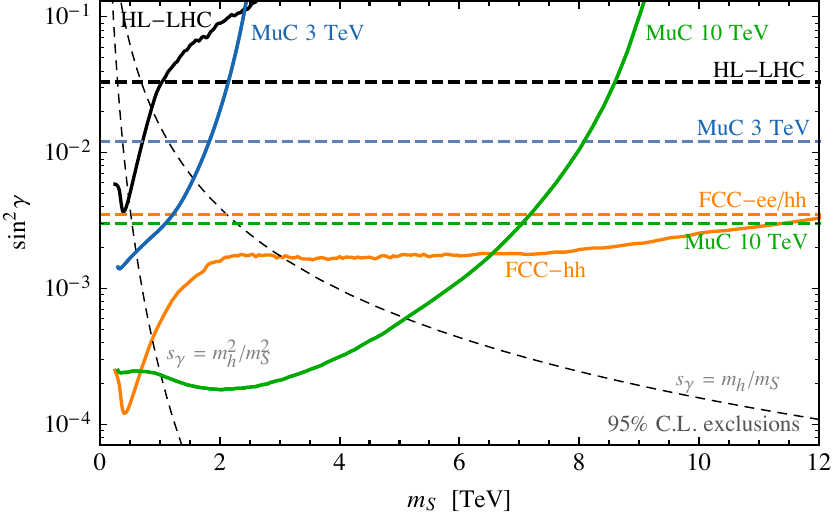}
\caption{\label{fig:singletmu} Constraints on the mixing angle as a function of the additional heavy scalar mass for a model with a simple singlet extension, taken from \cite{Accettura:2023ked}. Shown are direct {\sl (solid)} and indirect {\sl (slashed)} constraints for various future collider options. Direct constraints at the muon collider stem from VBF-type production of the heavy scalars. See original references for further details. } 
        \end{center}
    \end{figure}

As a last example, we consider the production of scalars in a 2HDM scenario, here taken in the decoupling limit where $\cos\lb \be-\al\rb\,\approx\,0$. In \cite{Accettura:2023ked} (based on work presented in \cite{Han:2021udl}), several production cross section plots have been presented for either pair-production or single production of the new scalars, the latter in association with a fermion pair, for direct production or VBF-type modes (similar cross sections have been presented in \cite{Kalinowski:2020rmb} for a model with a dark matter candidate.). We display these predictions in figure~\ref{fig:muon2hdm} as a function of the degenerate additional scalar masses $m_\Phi$, and with $\tan\be\,=\,1$.

    \begin{figure}[h!]
        \begin{center}
        \includegraphics[width=0.45\textwidth]{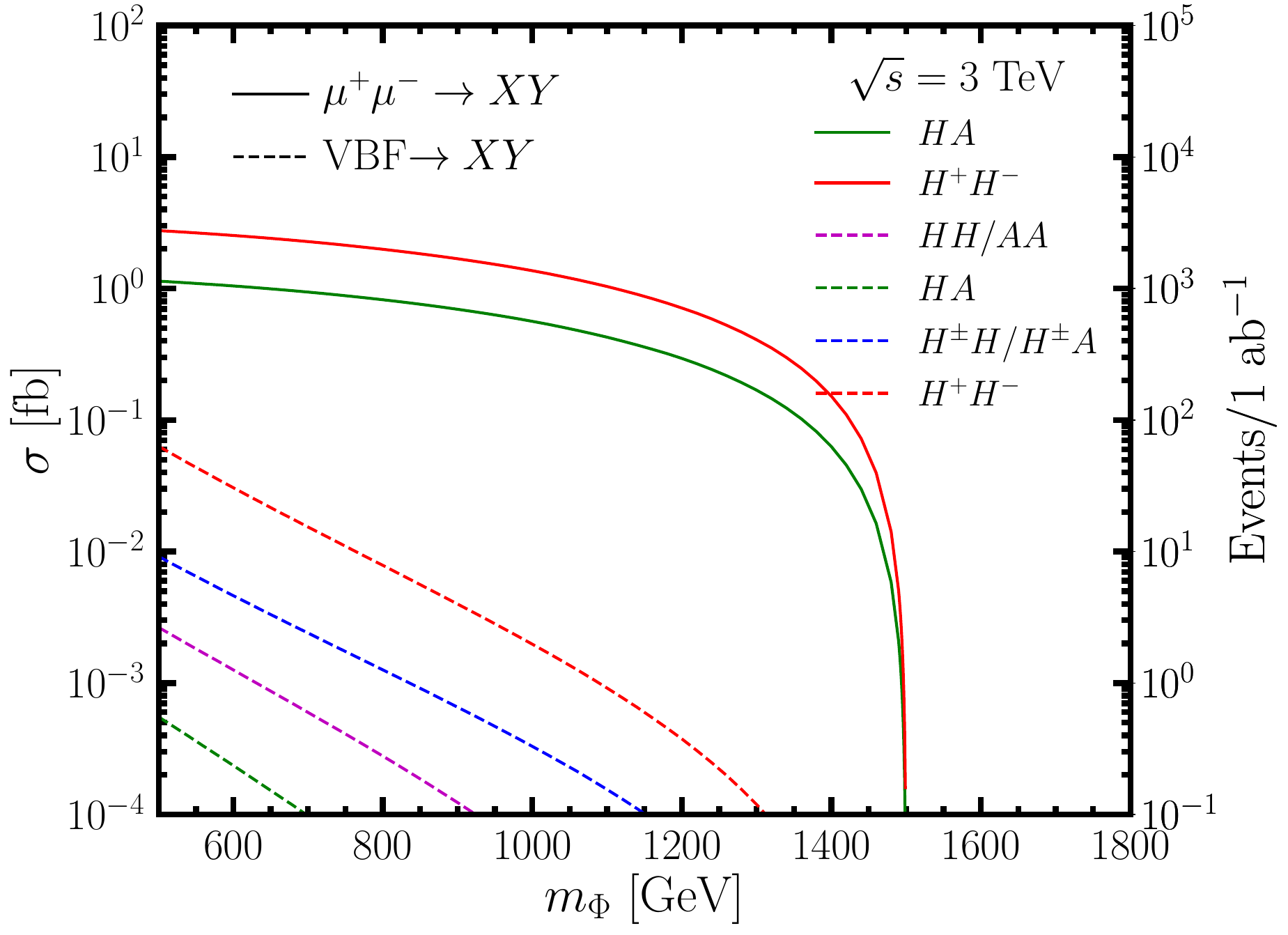}
         \includegraphics[width=0.45\textwidth]{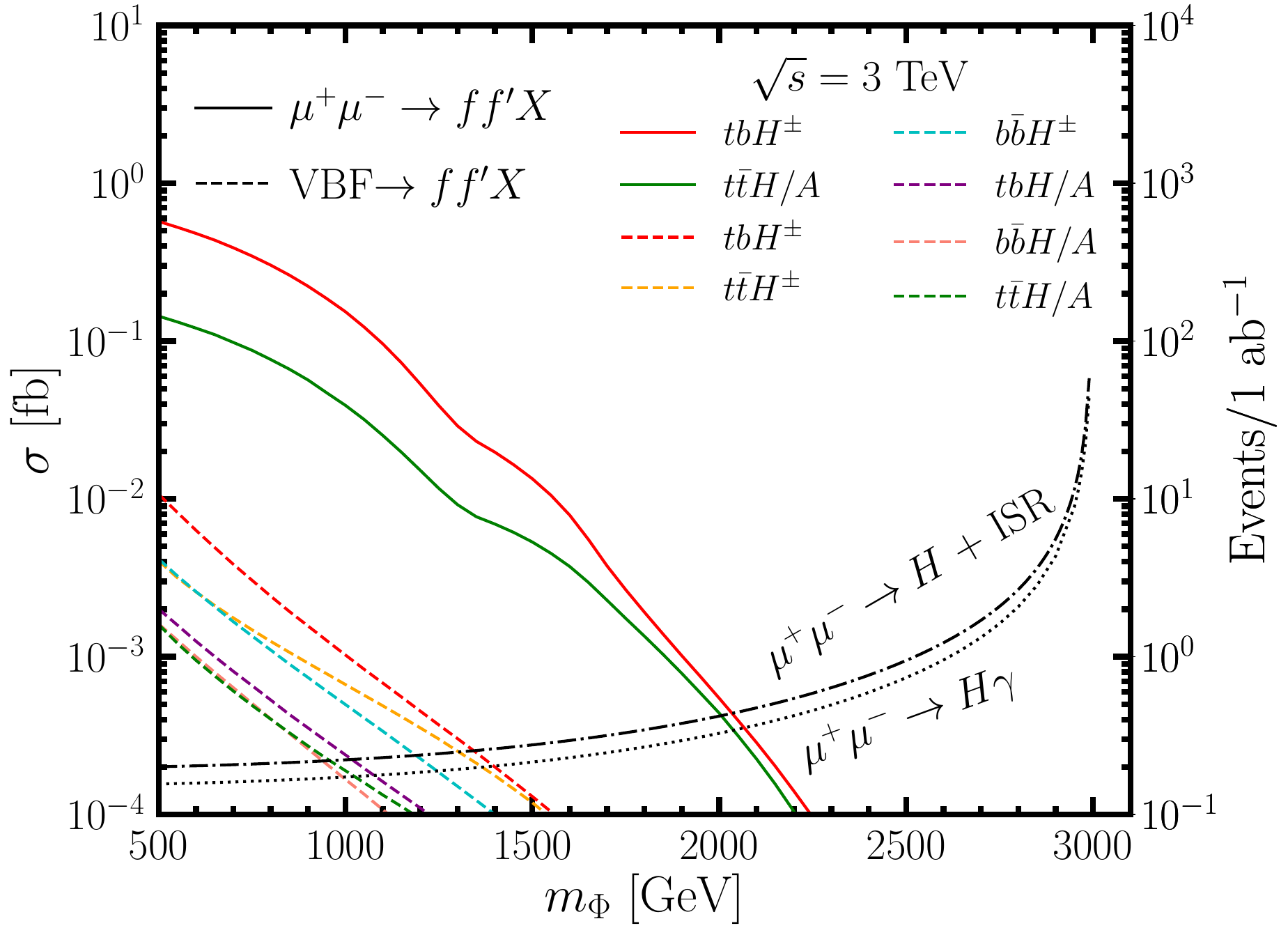}
        \caption{\label{fig:muon2hdm} Production cross sections at a 3 \TeV center-of-mass energy muon collider in a 2HDM, with degenerate additional scalar masses $m_\Phi$ and for $\cos\lb  \be-\al\rb\,\approx\,0$, $\tan\be\,=\,1$, for both direct production and VBF-type production modes. {\sl Left:} pair-production of the additional scalars. {\sl Right:} associate production mode including a di-fermion pair. Plot is taken from \cite{Accettura:2023ked}.}
        \end{center}
    \end{figure}

A detailed study for a model with an extended scalar sector containing a dark matter candidate at the 10 \TeV~ muon collider can e.g. be found in \cite{Braathen:2024ckk}.

}

\section{Connection to Cosmology/ electroweak phase transitions/ gravitational waves}\label{sec:cosmo}
{Today, the story we tell about the universe, is that first there was the Big Bang, and for a long time the universe was extremely hot, dense and symmetric. One could say that the symmetries of the SM were exact and the particles were massless, except for thermal masses. In fact, the originally massless particles in the unbroken phase, behave as if they had acquired a mass via the interactions with the plasma. 
At about approximately $10^{-12}$ seconds after the Big Bang (a temperature of 160 GeV) bubbles of the true vacuum started to form, nucleated  and finally gave rise to the universe we live in today. Therefore, the history of the universe is to a great extent the history of the Higgs potential. This shows how important extensions of the scalar sector can be.

Let us start with the vacuum problem at the present temperature. In quantum field theory, the vacuum can change the properties of a model. In fact, if a given field has a non-zero vacuum expectation value, a symmetry may be broken, and the corresponding quantum number is not conserved. If a VEV breaks the electromagnetic $U(1)$, electric charge will no longer be conserved. Although the SM conserves CP and electric charge in the Higgs potential due to gauge invariance, many extensions of the SM have the possibility of breaking both CP and charge. At zero temperature charge breaking should not be allowed and constraints on the model parameters are imposed to avoid such scenarios. On the other hand, at non-zero temperature, charge breaking VEVs could give rise to unexpected steps during the evolution of the universe~\cite{Ginzburg:2009dp, Barroso:2012mj}.

One of the problems of having just one Higgs field is the generation of the 
observed baryon asymmetry of the universe. Considering that the explanation
of this asymmetry lies in electroweak baryogenesis (EWBG) \cite{Kuzmin:1985mm,Cohen:1990it,Cohen:1993nk,Quiros:1994dr,Rubakov:1996vz,Funakubo:1996dw,Trodden:1998ym,Bernreuther:2002uj,Morrissey:2012db}, the three Sakharov conditions~\cite{Sakharov:1967dj}: Baryon number violation, C-symmetry and CP-symmetry violation, interactions out of thermal equilibrium, have to be fulfilled. The type of phase transition from the unbroken (zero VEV) to the broken phase is regulated by the Higgs potential, and the value of its parameters. Because the Higgs mass was measured to be 125 GeV at the LHC, in the scenario of a SM Higgs potential the transition would be a smooth cross-over \cite{Kajantie:1996mn,Csikor:1998eu} in contradiction with the requirement of a strong first-order electroweak phase transition
(SFOEWPT) needed for EWBG.
Extensions of the SM can solve the problem because the transition from the false to the true vacuum depends on the potential. A potential with extra scalars opens the possibility of a SFOWPT in some of its parameter space. This in turn can be a guide for searches for new scalar particles both at the LHC and at future colliders

If one assumes that the Higgs potential at zero temperature is a result of a SFOWPT in the Higgs vacuum that occurred in the early universe, a stochastic
gravitational wave (GW) background could appear as a sign of that transition and, under particular circumstances, could be detected in a not too distant future~\cite{Witten:1984rs, Kamionkowski:1993fg}. 
With the first observation of GWs by the LIGO Collaboration in 2016~\cite{LIGOScientific:2016aoc}, a new era of multi-messenger
astronomy started. Future GWs experiments such as LISA could give us further insight on the shape of the Higgs potential.

One question one should ask is again what is the minimal extension of the SM that is sufficient to trigger a strong enough FOPT. Again, the addition of a real gauge singlet is enough for EWBG~\cite{Espinosa:2007qk}. The possible detection of GWs in these simple extensions were first discussed in~\cite{Ashoorioon:2009nf}. Other extensions like adding an arbitrary number of singlets~\cite{Kakizaki:2015wua} or 2HDM scenarios~\cite{Dorsch:2013wja, Basler:2016obg} were also considered in the literature. These studies focus on how the parameters of the potential influence the duration and energy release of the phase transition and the properties of the primordial GWs, namely its peak amplitude and frequency.

It is also worth mentioning that different realizations of spontaneous symmetry breaking in a given model may lead to different strengths in the GWs signal.  
In fact, taking as simple example the complex extension of the SM, the different patterns of symmetry breaking (like having or not a DM candidate) may lead to substantial differences in the GWs spectrum~\cite{Freitas:2021yng}.
For recent reviews on phase transitions and gravitational waves see~\cite{Caprini:2019egz, Hindmarsh:2020hop, Athron:2023xlk}.

}

\section {Public tools}\label{sec:tools}

{
In this section, we present a list of some of the currently available public tools for several types of calculations in extended scalar sectors,
that are widely used within the community. We do not claim completeness and refer the reader to the respective manuals for further information on each program.

Any extension of the SM has to be tested against the constraints described in the previous sections. To do this one starts by writing the Lagrangian of the new model and determine the corresponding Feynman rules. The codes used by the community to go from the Lagrangian to the Feynman rules are mainly FeynRules~\cite{Alloul:2013bka}, LanHEP~\cite{Semenov:2014rea} and Sarah~\cite{Staub:2009bi}. 

In many models the constraints have to be checked with no help from ready-to-use codes. However, some codes can be applied to a large number of models.  HiggsTools~\cite{Bahl:2022igd} allows for comparison of models with extended scalar sectors to collider measurements and searches. It comprises the earlier developed packages HiggsBounds~\cite{Bechtle:2008jh,Bechtle:2011sb,Bechtle:2013wla,Bechtle:2020pkv}, allowing for comparison with direct search bounds from colliders, as well as HiggsSignals~\cite{Bechtle:2013xfa,Bechtle:2020uwn}, that renders information on the compatibility with 125 \GeV~ resonance measurements by the LHC experiments. The tool comes with some sample programs, and allows for several input format options.
Other codes like SuperIso~\cite{Mahmoudi:2007vz, Arbey:2018msw} can be used to constrain the models using flavor low energy observables. Other codes can be interfaces with these tools.

The next step is to apply the constraints to the model under scrutiny. This can be done by private codes but there are some available tools for a number of models. There are codes for the 2HDM, 2HDMC~\cite{Eriksson:2009ws,Eriksson:2010zzb}, and thdmtools~\cite{Biekotter:2023eil}.
ScannerS~\cite{Coimbra:2013qq,Muhlleitner:2020wwk} is a tool that allows for parameter scans for a large number of models with extended scalar sectors, including models with dark matter candidates. There is also a scanning tool for the Georgi-Machacek model~\cite{Hartling:2014xma}. Some of these codes have the theoretical constraints for the models inbuilt.

The decays of the several scalars in a new theory are available in 2HDMC and in HDecay~\cite{Djouadi:1997yw,Djouadi:2018xqq} for the 2HDM. Extensions of HDecay for models with an extra singlet, two doublets and one singlet and for the complex 2HDM are available in sHDECAY~\cite{Costa:2015llh}, N2HDECAY~\cite{Engeln:2018mbg}, C2HDECAY~\cite{Fontes:2017zfn}, respectively.

There are tools for the calculation of cross sections and branching ratios. Two very general codes in which the user can provide the Feynman rules are MadGraph5\_aMC@NLO~\cite{Alwall:2014hca, Frederix:2018nkq} and Calchep~\cite{Belyaev:2012qa}. MadGraph5\_aMC@NLO also include QCD corrections for some scenarios and can be used to generate events that after interfaced with other codes such as Pythia~\cite{Bierlich:2022pfr}. The detector simulator DELPHES~\cite{deFavereau:2013fsa} is widely used to perform experimental analyses.

In many extensions of the SM, the strength  of the Yukawa couplings relative to the SM can change. Therefore, also quantum corrections in single Higgs and double Higgs production via gluon fusion may change, because they proceed via quark loops. There are codes that handle this difference for single Higgs production, HIGLU~\cite{Spira:1995mt} and SusHi~\cite{Harlander:2012pb}, and for double Higgs production, HPAIR~\cite{Plehn:1996wb, Dawson:1998py}. Precise predictions for the trilinear Higgs couplings in the SM and BSM models, including higher-order corrections, are available from the public code anyH3~\cite{Bahl:2023eau}.

In situations where loop corrections are important but not available in public codes, a set of programs can be used for these calculations. Once the Feynman rules for the model are generated, FeynArts~\cite{Hahn:2000kx} can be used to generate the amplitude for the process at 1-loop. The calculation of the squared amplitude in terms of Passarino-Veltman (PV) functions is performed with FeynCalc~\cite{Mertig:1990an, Shtabovenko:2023idz}. Finally the PV functions are evaluated either by LoopTools~\cite{vanOldenborgh:1989wn, Hahn:1998yk} or by Collier~\cite{Denner:2016kdg}.

When the model includes DM candidates micrOMEGAs~\cite{Belanger:2001fz, Belanger:2006is, Alguero:2023zol} is a tool that calculates dark matter related observables such as DM relic density, direct and indirect detection cross sections, including detailed information about contributing channels and the 125 GeV Higgs invisible width. It contains a number of built-in models and allows for further model implementation via CalcHEP. There are other codes that calculate DM observables for BSM extensions, in particular the DM relic density, such as {\tt SuperIso Relic} \cite{Arbey:2018msw}, {\tt DarkSUSY} \cite{Bringmann:2018lay}, {\tt RelExt} \cite{Capucha:2025iml}, or {\tt MadDM} \cite{Backovic:2013dpa, Ambrogi:2018jqj}.

Finally there are codes dedicated to study phase transitions, CosmoTransitions~\cite{Wainwright:2011kj}, and phase transition and gravitational waves {\tt BSMPT}~\cite{Basler:2018cwe, Basler:2024aaf}. There are several codes on the market which trace the minima of involved scalar potentials and some of them also provide bounce solutions and other interesting features, \texttt{Vevacious}~\cite{Camargo-Molina:2013qva}, \texttt{VevaciousPlusPlus}~\cite{Camargo-Molina:2014pwa}, \texttt{AnyBubble}~\cite{Masoumi:2017trx}, \texttt{EVADE}~\cite{Hollik:2018wrr, Ferreira:2019iqb}, \texttt{BubbleProfiler}~ \cite{Athron:2019nbd, Akula:2016gpl}, 
\texttt{PhaseTracer} \cite{Athron:2020sbe} (that can be linked to potentials
  implemented in {\tt FlexibleSUSY}~\cite{Athron:2014yba, Athron:2017fvs} and {\tt
  BSMPT}), {\tt FlexibleSUSY}~\cite{Athron:2014yba, Athron:2017fvs}, ~\texttt{SimpleBounce} \cite{Sato:2019wpo, Sato:2019axv}, \texttt{FindBounce}~\cite{Guada:2018jek, Guada:2020xnz}, \texttt{OptiBounce}~\cite{Bardsley:2021lmq} and PT2GWFinder~\cite{Brdar:2025gyo}. Strong first-order phase transitions (SFOPT) during the evolution of the Higgs potential in the early universe not only allow for the dynamical generation of the observed matter-antimatter asymmetry, they can also source a stochastic gravitational wave background possibly detectable with future space-based gravitational waves interferometers. Different models have different behavior regarding the pattern of phase transitions and the production of gravitational waves.

Many publications use their own partially private codes which however partially contain or interface to the above tools.

\section{Conclusions}
\label{sec:conclusions}

{
As always happened throughout history we too have, at the moment this entry to the encyclopedia is being written, several outstanding physics issues to deal with. In fact, it doesn't matter when this is being written because it applies to every moment in time - there are always new and strange facts in science waiting for some explanation, or at least some mathematical description. 
This is the time of dark matter, the time of matter winning over anti-matter, the time of dark energy, of the neutrino masses, among others, including a number of issues considered theoretical problems like the so-called hierarchy problem, the large difference between the Planck and the electroweak scale, and the difference in masses of the known fermions. From all these issues, extensions of the scalar sector of the Standard Model, have as a primary goal the resolution of the dark matter, baryon asymmetry and neutrino mass problems. 

This discussion on extended scalar sectors starts, after the introduction, with the theoretical and experimental constraints on new physics models. The proposed models have to be probed and the theoretical constraints remove the parameter space that is considered nonphysical or at least not suitable for the use of perturbation theory in the calculations. The experimental constraints eliminate the parameter space that is not compatible with the current experimental findings.

The simplest extensions of the scalar sector of the SM are then discussed. We start with the real scalar singlet, followed by the addition of one more doublet. These two models give rise to new particles that are being searched for at the Large Hadron Collider. Dark matter and the possibility of having a first-order phase transition can be achieved with singlets only but CP-violation needs at least one extra doublet, unless extra fermions are also added. Singlet extensions also improve the stability of the SM with points in parameter space that are stable and not meta-stable. More general models may tune the parameters needed for complete agreement with the dark matter and baryon asymmetry experimental results.

The next section is devoted to the construction of DM models with extended scalar sectors. The principle is that of a portal between a visible and a dark sector. There are unbroken symmetries to stabilize dark matter. CP-violation of a dark sector is also discussed. 

The LHC is probing a large number of new physics models including the ones discussed here. The current experimental status is presented taking just a few models and searches as examples, as well as the reach of possible future collider options. Except for the High Luminosity stage of the LHC, which is already approved, all other options are being discussed by the community in terms of the different physics cases. 
The following section focuses on a very brief discussion on how extended sectors connect to cosmology, to electroweak phase transitions and also gravitational waves. Finally, we end this work with a list of the most used public tools.

The inclusion of a spin zero field in the SM was motivated theoretically due to the Brout-Englert-Higgs mechanism and the need to give mass to all particles without breaking the SM symmetries explicitly. Hence, the search for the Higgs boson was one of main goals of LEP, the Tevatron and the LHC. Now that a 125 GeV scalar compatible with SM prediction was discovered, we again could have new spin zero particles that would solve outstanding issues of the SM. Although the theoretical motivation is not as strong, there are good reasons to keep search for new scalars both at the LHC and at future colliders.

}


\begin{ack}[Acknowledgments]%
We thank the editors of this encyclopedia for inviting us to submit this contribution. TR has received support from grant number HRZZ-IP-2022-10-2520 from the Croatian Science Foundation (HRZZ). RS is partially supported by the Portuguese Foundation for Science and Technology (FCT) under CFTC: UID/00618/2025
and through the PRR (Recovery and Resilience
Plan), within the scope of the investment “RE-C06-i06 - Science Plus Capacity Building”, measure “RE-C06-i06.m02 - Reinforcement of financing for International Partnerships in Science,
Technology and Innovation of the PRR”, under the project with the reference 2024.03328.CERN.
\end{ack}


\bibliographystyle{Numbered-Style} 
\bibliography{reference}

\end{document}